%% file: main.tex
\documentclass[letterpaper,twocolumn,10pt]{article}
\usepackage[svgnames]{xcolor}
\usepackage{usenix-2020-09}

\usepackage{hyperref}
\usepackage{pifont}
\usepackage{spverbatim}

\usepackage{multirow}
\usepackage[normalem]{ulem}
\useunder{\uline}{\ul}{}
\usepackage{subfigure}
\usepackage[ruled,vlined,linesnumbered]{algorithm2e}
\SetKwComment{Comment}{$\triangleright$\ }{}
\usepackage{graphicx}
\usepackage{caption}
\usepackage[skip=1pt]{caption}
\usepackage{float}
\usepackage[flushleft]{threeparttable}
\usepackage{array,booktabs,makecell}
\usepackage{graphicx}
\usepackage{enumitem}
\usepackage[framemethod=default]{mdframed}
\usepackage{mathtools,lipsum}

\usepackage{cuted}
\usepackage[toc,page]{appendix}
\usepackage{balance}
\usepackage{footmisc}
\usepackage{tabularx}
\usepackage{tikz}
\usepackage{xcolor}
\usepackage{multirow}
\usepackage[many]{tcolorbox}
\usepackage{subcaption}
\usepackage[export]{adjustbox}
\usepackage{adjustbox}

\usepackage{colortbl}%

\newcommand{\graycell}{\cellcolor[HTML]{EFEFEF}}

\usepackage{tikz}

\definecolor{darkred}{HTML}{860000}
\definecolor{darkteal}{HTML}{005959}
\definecolor{darkpurple}{HTML}{590059}
\definecolor{darkgrey}{HTML}{434343}

\newtcolorbox{mybox}[2][]{text width=0.95\linewidth,fontupper=\normalsize,
fonttitle=\bfseries\sffamily\scriptsize, colbacktitle=darkgrey,enhanced,
attach boxed title to top left={yshift=-2mm,xshift=3mm},
boxed title style={sharp corners},top=4pt,bottom=2pt,left=2pt,right=2pt,
  title=#2,colback=white}

\newcommand{\haoran}[1]{\textbf{\textcolor{purple}{[Haoran: #1]}}}
\newcommand{\tool}{\textsc{Oedipus}}

\usepackage{listings}

\def\BibTeX{{\rm B\kern-.05em{\sc i\kern-.025em b}\kern-.08emT\kern-.1667em\lower.7ex\hbox{E}\kern-.125emX}}

\begin{document}
\title{\tool{}: LLM-enchanced Reasoning CAPTCHA Solver}

\newcommand{\mkntu}[0]{{{$$}}}

\author{
    Gelei Deng\mkntu\textsuperscript{\textsection},
    Haoran Ou\mkntu\textsuperscript{\textsection},
    Yi Liu\mkntu,
    Jie Zhang\mkntu,
    Tianwei Zhang\mkntu, 
    Yang Liu\mkntu, 
    \small
    \\ \\
    \mkntu Nanyang Technological University 
    \\
}

\maketitle
\thispagestyle{plain}
\pagestyle{plain}

\begin{abstract}
CAPTCHAs have become a ubiquitous tool in safeguarding applications from automated bots. Over time, the arms race between CAPTCHA development and evasion techniques has led to increasingly sophisticated and diverse designs. The latest iteration, reasoning CAPTCHAs, exploits tasks that are intuitively simple for humans but challenging for conventional AI technologies, thereby enhancing security measures.

Driven by the evolving AI capabilities, particularly the advancements in Large Language Models (LLMs), we investigate the potential of multimodal LLMs to solve modern reasoning CAPTCHAs. Our empirical analysis reveals that, despite their advanced reasoning capabilities, LLMs struggle to solve these CAPTCHAs effectively. In response, we introduce \tool{}, an innovative end-to-end framework for automated reasoning CAPTCHA solving. Central to this framework is a novel strategy that dissects the complex and human-easy-AI-hard tasks into a sequence of simpler and AI-easy steps. This is achieved through the development of a Domain Specific Language (DSL) for CAPTCHAs that guides LLMs in generating actionable sub-steps for each CAPTCHA challenge. The DSL is customized to ensure that each unit operation is a highly solvable subtask revealed in our previous empirical study. These sub-steps are then tackled sequentially using the Chain-of-Thought (CoT) methodology. 

Our evaluation shows that \tool{} effectively resolves the studied CAPTCHAs, achieving an average success rate of 63.5\%. Remarkably, it also shows adaptability to the most recent CAPTCHA designs introduced in late 2023, which are not included in our initial study. This prompts a discussion on future strategies for designing reasoning CAPTCHAs that can effectively counter advanced AI solutions.

\end{abstract}

\input{Tex/1-Introduction}

\input{Tex/2-Background}
\input{Tex/3-Study}

\input{Tex/4-Methodology}
\input{Tex/5-Evaluation}

\input{Tex/6-Discussion}

\input{Tex/7-Conclusion}

\clearpage
\bibliographystyle{IEEEtran}

\bibliography{sample-base}

\input{Tex/Appendix}

\end{document}

%% file: Tex/1-Introduction.tex
\begingroup\renewcommand\thefootnote{\textsection}
\footnotetext{Equal Contribution}
\endgroup

\section{Introduction}\label{sec:intro}

The pervasive threat posed by automated bots has necessitated robust countermeasures to safeguard the integrity and functionality of various applications, leading to the development and widely application of CAPTCHAs~\cite{cap} (Completely Automated Public Turing test to tell Computers and Humans Apart). Ingeniously tapping into the distinct cognitive capabilities of humans versus the computational limitations of machines, CAPTCHAs present tasks that are trivial for humans but substantially challenging for automated systems. This delineation exemplifies Moravec's paradox~\cite{moravec}, which suggests that activities requiring minimal human thought are disproportionately difficult for artificial intelligence (AI) to replicate. By designing CAPTCHAs around this concept—creating tasks that are simple for humans but difficult for AI—these tests act as a formidable barrier against automated intrusions.

However, as AI technology is evolving, the effectiveness of traditional CAPTCHAs has diminished. Initial CAPTCHA challenges, such as ReCAPTCHA~\cite{ReCAPTCHA} and hCaptcha~\cite{hCaptcha}, focused on straightforward tasks like text recognition or basic image identification. These tasks leverage the visual and cognitive abilities that are innate to humans but were initially difficult for computer algorithms to replicate. With the advance of computer vision (CV)~\cite{ye2018yet} and machine learning~\cite{bursztein2014end} techniques, these early CAPTCHAs became increasingly vulnerable to automated solving~\cite{jin2023secure, hossen2021object}.

Such rapid development necessitates the creation of more complex CAPTCHA types, notably the reasoning CAPTCHAs~\cite{wang2018captcha,gao2021research,wang2023extended}. These CAPTCHAs mark a significant departure from earlier approaches, relying less on object recognition, a tractable problem for advanced CV techniques\cite{zong2022detrs}, but more on tasks requiring logical reasoning, problem-solving, and interpretation of complex instructions. These activities demand a level of cognitive engagement that current AI solutions struggle to provide, making reasoning CAPTCHAs a more robust defense mechanism. Despite the advancements in AI, reasoning CAPTCHAs continue to pose a significant challenge to automated solvers, leading to their adoption by many popular online platforms such as LinkedIn~\cite{linkedin}, TikTok~\cite{tiktok} and Twitter~\cite{twitter}.

The landscape of AI technology, particularly with the evolution of Large Language Models (LLMs)~\cite{zhao2023survey}, has been marked by significant advancements, notably in reasoning capabilities and multimodal processing. These developments have paved the way for novel approaches to tackling reasoning CAPTCHAs, tasks traditionally considered challenging for AI due to their reliance on human-like cognitive processes. To evaluate the efficacy of these advanced AI models in the context of reasoning CAPTCHA solving, this paper embarks on an empirical investigation, which engages two leading-edge multimodal LLMs (GPT-4V(ision)~\cite{gpt4} and Gemini~\cite{gemini}), utilizing zero-shot prompting~\cite{liu2023pre} and the Chain-of-Thoughts (CoT) strategy~\cite{wei2022chain} as primary methodologies (Section \ref{sec:study}). Our objective is to explore the extent of their capabilities and delineate the boundaries within which these models operate when confronted with various reasoning CAPTCHAs.

The outcomes of our investigation reveal that, despite the unprecedented capabilities of LLMs, they currently fall short in effectively solving reasoning CAPTCHAs. Our analysis yields four key insights: (1) LLMs exhibit a comprehensive understanding of CAPTCHA tasks, including the challenges posed and objectives to be achieved. (2) Unexpectedly, LLMs are capable of deconstructing complex reasoning tasks into simpler, manageable steps and addressing each through the CoT strategy similar to human being. However, the efficacy of this approach is contingent upon the models' success in accurately completing every step in the sequence; failure of any step inevitably results in the failure of the entire task. (3) The primary reason of models' failure in CAPTCHA solving is their limited capabilities of recognizing objects. While they can recognize and attribute characteristics to singular objects within CAPTCHAs, this ability significantly diminishes when tasked with simultaneously discerning attributes of multiple objects, which is often required in reasoning CAPTCHAs. (4) Our study indicates that LLMs face challenges in executing multiple reasoning steps within a single prompt, with a notable increase in errors and hallucinations as the number of required reasoning steps escalates, leading to the unsuccessful resolution of CAPTCHA challenges.

This empirical study prompts us to consider the possibility of breaking down an \textit{AI-hard} reasoning CAPTCHA challenge into a series of \textit{AI-easy} tasks that can be more readily solved by LLMs. More specifically, we expect a strategy that can break a given reasoning CAPTCHA into a series of detailed operations, such that every operation aligns with LLM capabilities as revealed by our empirical study. To this end, we propose our CAPTCHA Domain Specific Language  (DSL)~\cite{mernik2005and} to regulate this task breakdown process (Section \ref{sec:dsl}). We carefully design the operations and syntax of this CAPTCHA DSL, such that the syntax-correct CAPTCHA DSL scripts will only contain operations that are highly solvable by LLMs. 
We direct LLMs to generate challenge solutions in the CAPTCHA DSL scripts that adhere to these syntax principles. A local debugger aids in refining these solutions by identifying and correcting inaccuracies, thereby enabling a systematic approach to generating solutions for reasoning CAPTCHA challenges.

Building on this foundation, we present an end-to-end framework, \tool{}\footnote{Oedipus was a mythical Greek king who answered Sphinx's riddles correctly, and defeated this monster.}, that utilizes our CAPTCHA DSL to automate the solving of reasoning CAPTCHAs (Section \ref{sec:methodology}). The workflow initiates with the creation of a DSL script for a given CAPTCHA challenge, followed by its translation into natural language instructions that LLMs can understand and execute. By providing the natural language instructions together with the original CAPTCHA challenge, a multimodal LLM is able to solve it step by step. Our framework offers two profound advantages. First, compared to traditional deep learning-based CAPTCHA solving strategies~\cite{ye2018yet,noury2020deep}, \tool{} does not require any training process or collection of labeled training data, which saves great manual efforts in the domain of CAPTCHA solving. Second, \tool{} is adaptable to new CAPTCHA types as long as the unit operations required for solving them are covered by the CAPTCHA DSL.

The efficacy of \tool{} is validated through extensive evaluations and analysis. We deploy \tool{} on 4 types of reasoning CAPTCHAs designed by 2022 and commercially available online. The experimental results are promising, demonstrating that \tool{} achieves a success rate of 63.5\% in solving these CAPTCHAs on average, with a cost of lowest to 3.1 USD per 100 CAPTCHA solving. Among these CAPTCHAs, two types have never been solved by any existing solutions discussed in academia. Additionally, \tool{} has shown proficiency in resolving two newly developed CAPTCHAs from 2023 with an average success rate of 44.1\%, whose solutions are not used to guide the development of the CAPTCHA DSL generator component. This underscores the versatility of \tool{}: as long as the required operations are within the scope of our CAPTCHA DSL, \tool{} can consistently perform effectively without further training process.

In light of our findings, we propose three strategies for designing advanced CAPTCHAs that could potentially remain unsolvable by current LLMs. First, we suggest CAPTCHAs that demand complex reasoning chains beyond LLMs' current reach, pushing the limits of AI problem-solving. Second, we recommend employing adversarial examples \cite{goodfellow2014explaining} to exploit and confuse LLMs' object recognition abilities. Finally, we advocate for CAPTCHAs requiring an understanding of concepts or operations that lie outside the scope of existing LLMs, such as intricate real-world interactions. Acknowledging the rapid advancement of LLMs, these strategies aim to maintain CAPTCHAs as effective security measures by continuously adapting to and anticipating future AI developments, ensuring a dynamic balance between CAPTCHA complexity and AI solving capabilities.

\noindent\textbf{Ethical Declaration.} We emphasize that our research and the development of \tool{} have not been leveraged for any unethical activities or financial gain. We are acutely aware of the ethical implications of our work. Therefore, in compliance with ethical standards, we refrain from releasing a fully automated CAPTCHA  solving tool. Instead, we provide access to a partial solution that generates CAPTCHA  resolutions in natural language. This approach ensures that while we contribute to the advancement of AI in CAPTCHA  solving, we also maintain a strong commitment to ethical practices in AI research and development.
In the spirit of responsible disclosure, 
we have also proactively communicated our discoveries to the companies involved in CAPTCHA development.

%% file: Tex/2-Background.tex
\section{Background}\label{sec:background}

\subsection{CAPTCHAs  and CAPTCHA  Solver}




CAPTCHAs~\cite{captcha-background,captcha-survey} have evolved from simple text puzzles to complex challenges to distinguish humans from bots. This evolution responds to advancements in solver algorithms, transitioning from basic OCR~\cite{captcha-solver-1} to advanced machine learning models~\cite{captcha-solver-2,captcha-solver-3,captcha-solver-4}. As CAPTCHAs become more intricate, solvers adapt to creating a continuous arms race in web security. This cycle highlights the need for innovative approaches to security, suggesting a shift towards analyzing user behaviors and incorporating AI to maintain the balance between user accessibility and protection against automated threats.

\subsection{Reasoning CAPTCHAs}


The development of reasoning CAPTCHAs~\cite{gao2021research} represents a significant advancement in web security, shifting focus from visual recognition to cognitive challenges that require logical deduction and contextual understanding. This evolution counters automated solvers more effectively than traditional CAPTCHAs by demanding tasks that challenge standard neural networks. By exploiting the unique human capability for complex reasoning, reasoning CAPTCHAs provide a stronger defense against bots, marking a critical step forward in protecting online interactions.

\subsection{Large Language Models}



LLMs~\cite{llm-survey} have emerged as a groundbreaking advancement in AI, demonstrating remarkable abilities in understanding and generating natural language across a broad spectrum of applications. The evolution into multimodal LLMs, capable of processing and integrating various types of data such as text and images, underscores their potential to tackle complex, multi-dimensional problems. Notably, the reasoning capabilities inherent in LLMs position them as promising candidates for security tasks, such as penetration testing~\cite{deng2023pentestgpt} and protocol fuzzing~\cite{meng2024large}. This makes LLMs potentially suitable for reasoning CAPTCHAs solving. Although the application of LLMs in this context has yet to be thoroughly explored, their sophisticated advancements in handling complex reasoning tasks suggest a significant potential in overcoming the challenges presented by reasoning CAPTCHAs, highlighting a new frontier in the application of AI in web security.

%% file: Tex/3-Study.tex
\section{Empirical Study}\label{sec:study}

Previous studies~\cite{wang2023bot} have demonstrated the effectiveness of LLMs in solving traditional CAPTCHAs, such as those offered by ReCAPTCHA. However, the potential of LLMs in addressing the more complex reasoning CAPTCHAs remains unexplored. To bridge this gap, we embark on an empirical study aimed at understanding the capabilities of LLMs in solving reasoning CAPTCHAs. This investigation is structured around two pivotal research questions:

\begin{itemize}[leftmargin=*,itemsep=1pt,topsep=0pt,parsep=1pt]
\item\textbf{RQ1 (Categorization)}: What are the different types of reasoning tasks present in reasoning CAPTCHAs?

\item\textbf{RQ2 (Effectiveness)}: How effective are LLMs in accurately solving reasoning CAPTCHAs, and what factors influence their success rate?
\end{itemize}
We detail the empirical study results in the following. 


\subsection{CAPTCHA  Categorization (RQ1)}

\noindent\textbf{Dataset Construction.} To comprehensively address RQ1, our primary objective is to categorize existing reasoning CAPTCHAs. For this purpose, we embark on a large-scale enumeration of CAPTCHAs available online. Specifically, we review all CAPTCHAs referenced in recent research works~\cite{gao2021captcha, dinh2023recent, andrew2023captcha, kumar2022systematic}, with a focus on commercially available reasoning CAPTCHAs. It is important to note that our study excludes CAPTCHAs exclusively discussed in academic research, such as those in~\cite{291062}, due to the absence of readily available APIs and their untested effectiveness in real-world scenarios. Consequently, we have compiled a dataset comprising 5 types\footnote{The CAPTCHAs' name are given by the vendors.} of reasoning CAPTCHAs from \textcolor{blue}{3} different vendors, detailed in Table~\ref{tab:category}.

\begin{figure}[h]
	\centering
	\begin{minipage}[b]{\linewidth}
		\centering
            \subfigure[CAPTCHA 1 by Arkose Labs]{
            \begin{minipage}[b]{0.47\linewidth}
                \centering
                \includegraphics[width=\linewidth]{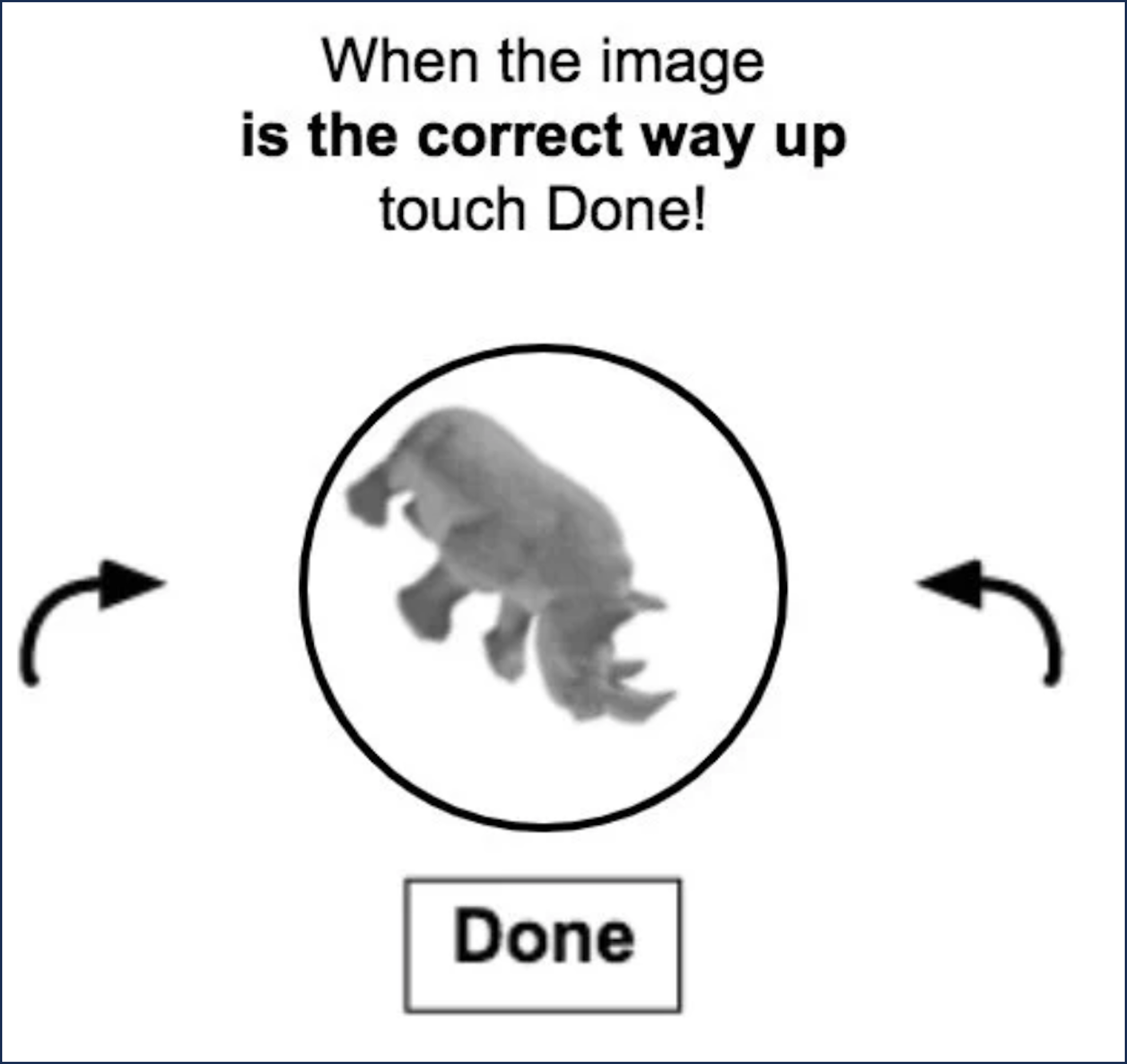}
                \vspace{-10pt}
                \label{fig:rotation1}
            \end{minipage}
            }
            \subfigure[CAPTCHA 2 by Arkose Labs]{
            \begin{minipage}[b]{0.47\linewidth}
                \centering
                \includegraphics[width=\linewidth]{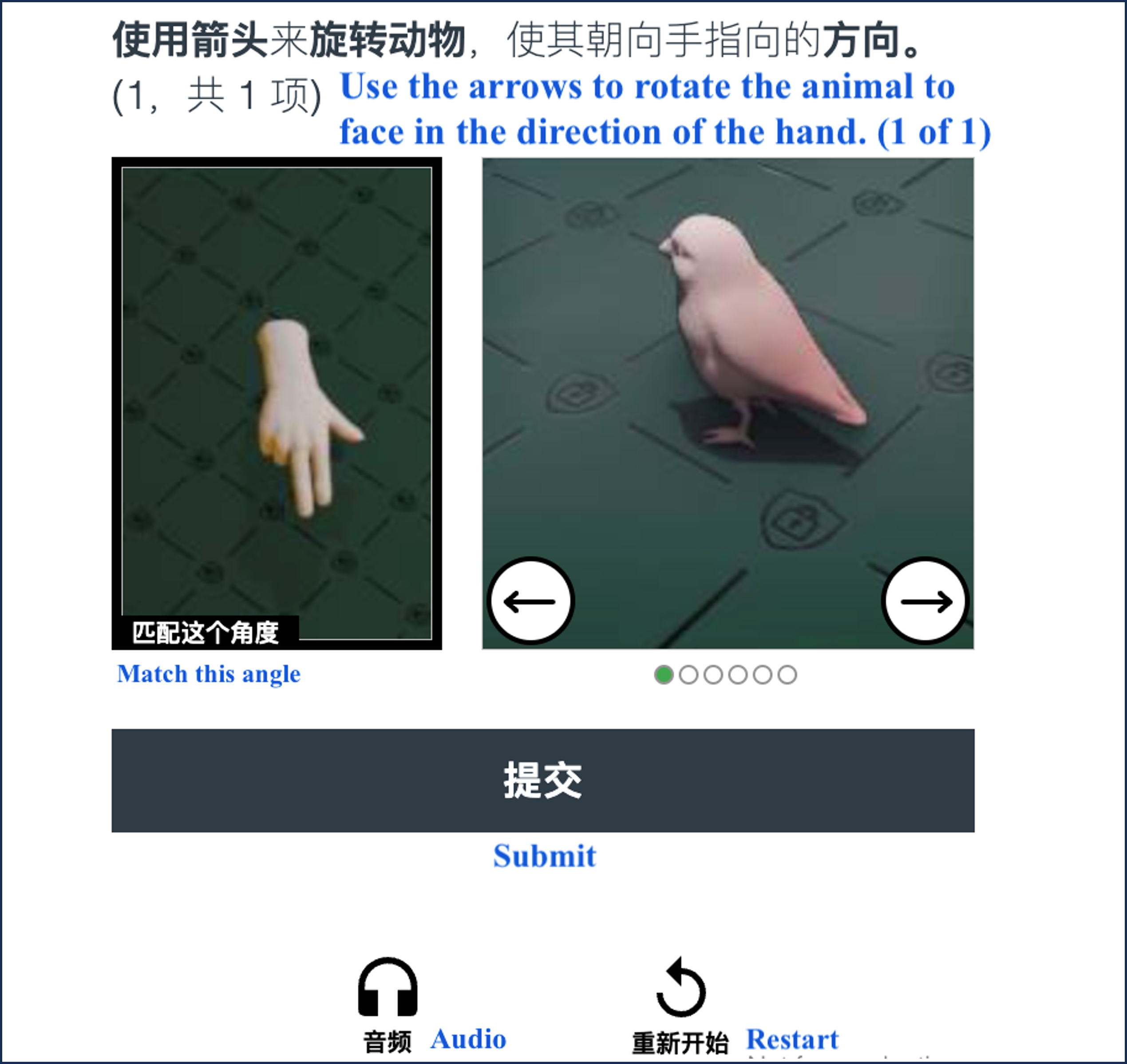}
                \vspace{-10pt}
                \label{fig:rotation2}
            \end{minipage}
            }
            \caption{Rotation CAPTCHA  examples.}
            \label{fig:rotation-CAPTCHA-examples}
	\end{minipage} 
        \hfill
	\begin{minipage}[b]{\linewidth}
		\centering
            \subfigure[CAPTCHA 1 by GeeTest]{
            \begin{minipage}[b]{0.47\linewidth}
                \centering
                \includegraphics[width=\linewidth]{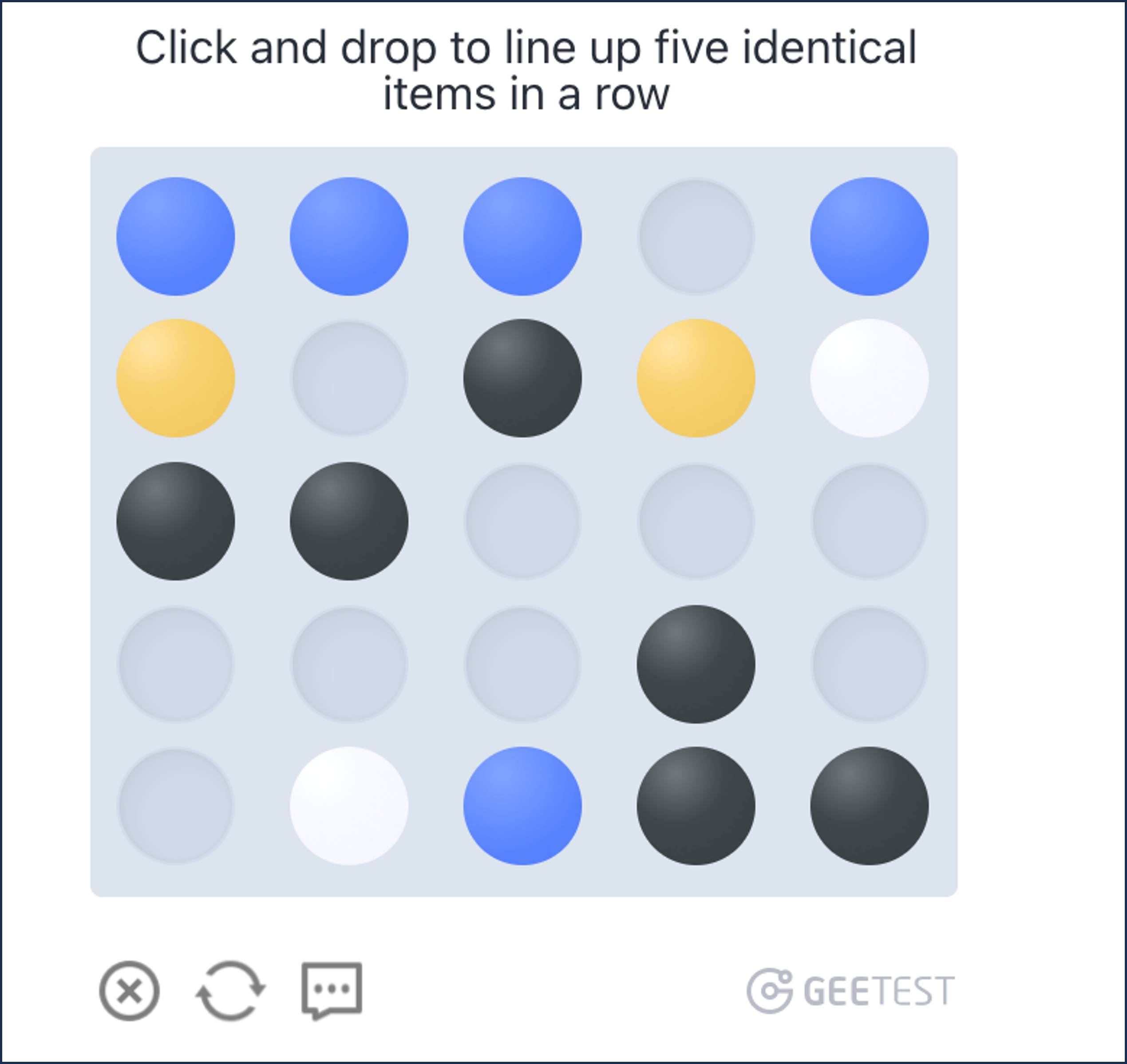}
                \vspace{-10pt}
                \label{fig:bingo1}
            \end{minipage}
            }
            \subfigure[CAPTCHA 2 by GeeTest]{
            \begin{minipage}[b]{0.47\linewidth}
                \centering
                \includegraphics[width=\linewidth]{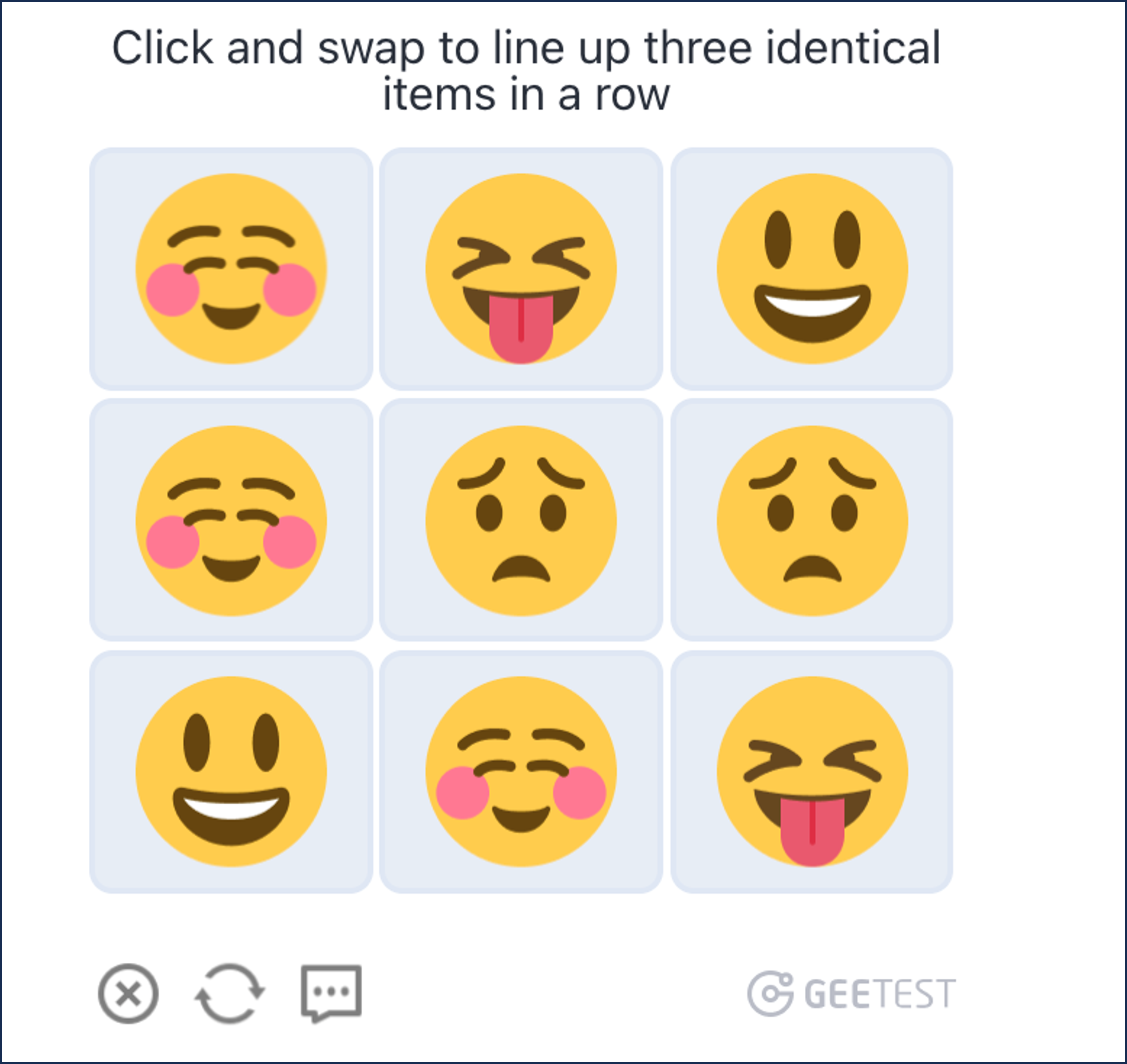}
                \vspace{-10pt}
                \label{fig:bingo2}
            \end{minipage}
            }
            \caption{Bingo CAPTCHA examples.}
            \label{fig:bingo-CAPTCHA -examples}
	\end{minipage}
        \hfill
        \begin{minipage}[b]{\linewidth}
		\centering
            \subfigure[CAPTCHA 1 by YiDun]{
            \begin{minipage}[b]{0.47\linewidth}
                \centering
                \includegraphics[width=\linewidth]{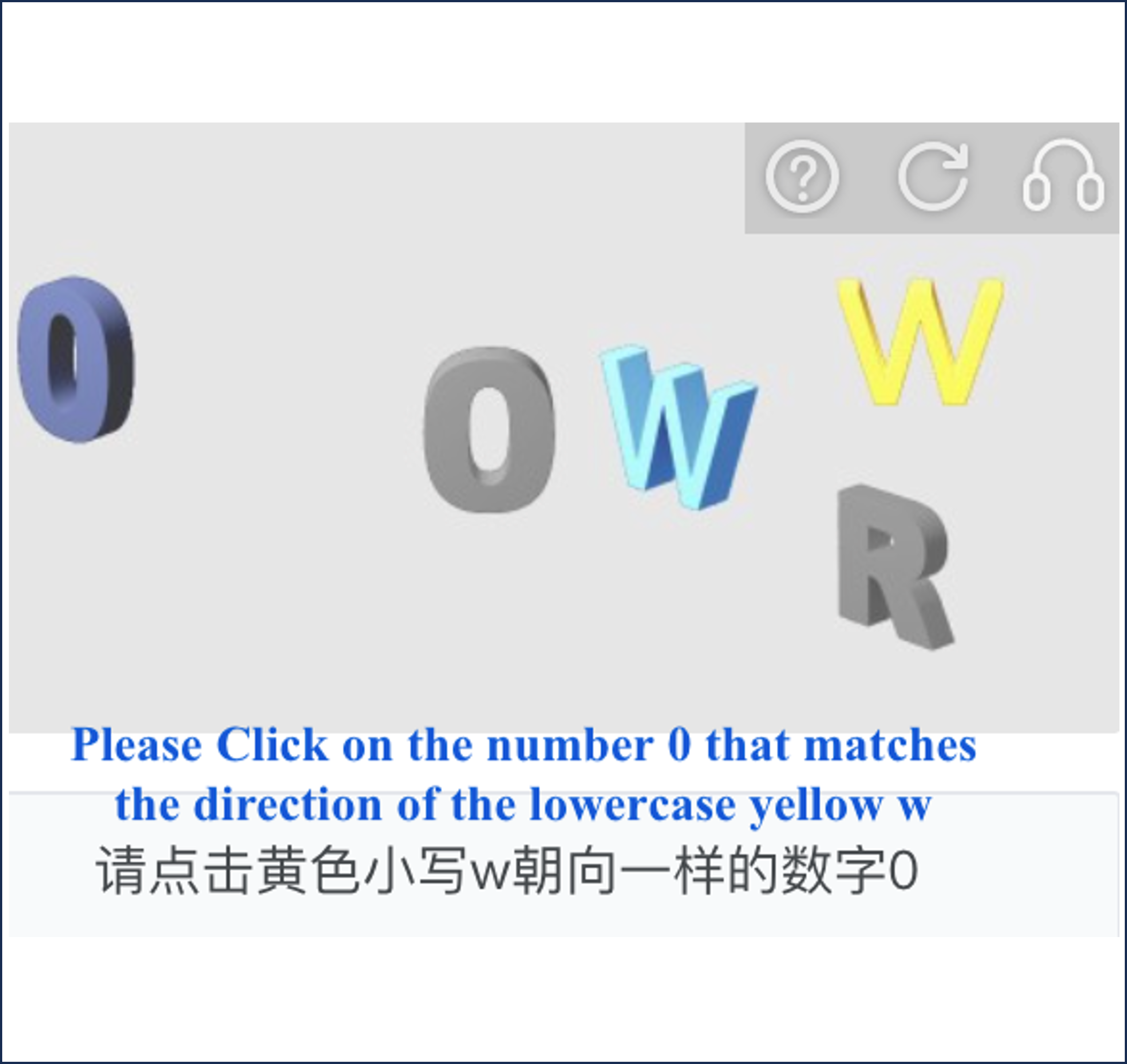}
                \vspace{-10pt}
                \label{fig:3d1}
            \end{minipage}
            }
            \subfigure[CAPTCHA 2 by GeeTest]{
            \begin{minipage}[b]{0.47\linewidth}
                \centering
                \includegraphics[width=\linewidth]{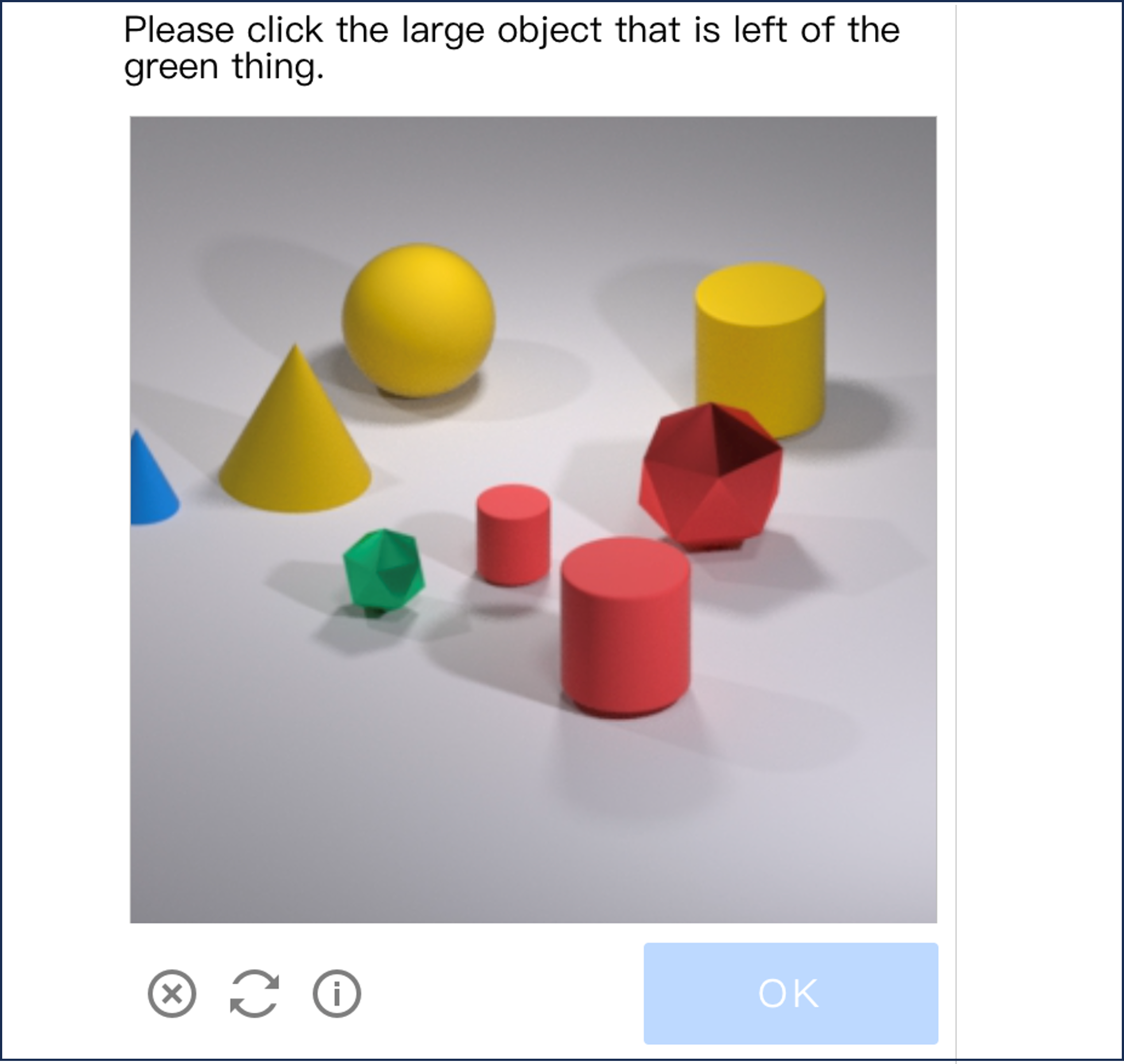}
                \vspace{-10pt}
                \label{fig:3d2}
            \end{minipage}
            }
            \caption{3D CAPTCHA examples.}
            \label{fig:3d-captcha-examples}
	\end{minipage}
 \vspace{-25pt}
\end{figure}

\noindent\textbf{CAPTCHA  Categorization.} In alignment with preceding works~\cite{gao2021captcha, andrew2023captcha}, our categorization of CAPTCHAs is fundamentally based on the reasoning tasks embedded in their designs. Ultimately, we identify 3 distinct categories of reasoning CAPTCHAs:

\begin{enumerate}[leftmargin=*,itemsep=1pt,topsep=0pt,parsep=1pt]
\item \textbf{Rotation CAPTCHAs.} Displayed in Figure~\ref{fig:rotation-CAPTCHA-examples}, rotation CAPTCHAs compel users to adjust an object's orientation to match that of a reference object. This type has evolved from the conventional rotation CAPTCHAs (seen in Figure~\ref{fig:rotation1}) described in~\cite{andrew2023captcha}, which merely required upward orientation of objects. The newer variants, as depicted in Figure~\ref{fig:rotation2}, introduce increased complexity by necessitating users to discern and align with the orientation of a reference item, such as a finger in the provided example.
\item \textbf{Bingo CAPTCHAs.} In Figure~\ref{fig:bingo-CAPTCHA -examples}, bingo CAPTCHAs present the challenge of identifying elements on a board and rearranging them to form a line of identical items. These CAPTCHAs vary greatly in terms of element types and manipulation rules, contingent on the provider. For instance, Figure~\ref{fig:bingo1} allows arbitrary swapping of any two items, while Figure~\ref{fig:bingo2} restricts users to only swap adjacent items, demonstrating the diversity in this category.
\item \textbf{3D Logical CAPTCHAs.} Illustrated in Figure~\ref{fig:3d-captcha-examples}, 3D Logical CAPTCHAs task users with selecting an object from a 3D space, based on intricate logical relationships involving attributes like shape, color, and orientation. For example, Figure~\ref{fig:3d1} challenges users to identify the number 0 that shares the same orientation as a yellow letter W; whereas Figure~\ref{fig:3d2} requires selecting the larger object positioned to the left of a green object.
\end{enumerate}

The unique set of challenges posed by each CAPTCHA category necessitates specialized solving strategies. By categorizing these CAPTCHAs, our goal is to gain a thorough understanding of the array of reasoning tasks they encompass and critically assess the proficiency of LLMs in tackling these diverse challenges.

\begin{table*}[ht]
  \centering
  \begin{tabular}{lllll}
  \toprule
\textbf{Name} & \textbf{Vendor} & \textbf{Category} & \textbf{\# of Samples} & \textbf{Description} \\ \hline
FunCAPTCHA1      &  Arkose Labs      & Rotation    &     10          &  Rotate the object into the indicated direction           \\
Gobang      & GeeTest      & Binto    &            10   & Line up five identical items in a row            \\
IconCrush      & GeeTest      & Bingo    &         10      &  Line up three identical items in a row           \\
Space CAPTCHA      & GeeTest      & Space    &      10         &  Based on the space relation click the indicated object           \\
Space Reasoning      & YiDun      & Space Reasoning    &    10           & Based on the logical relation click the indicated object       \\ \midrule
\end{tabular}
  \caption{Summary of collected CAPTCHAs.}
  \label{tab:category}
\end{table*}

There are two things that are worth noting. Firstly, there is a noticeable diversity among vendors in the types of reasoning CAPTCHAs developed, with little overlap in design. This finding is underscored by our discovery that CAPTCHA  companies have obtained patents for their unique designs, as can be verified at \cite{patent}.
Second, CAPTCHAs can be broadly classified into two distinct categories based on their design complexity and feasibility of exhaustively cataloging their variations. The first category is ``Limited Variability'' CAPTCHAs, such as rotation CAPTCHAs. These CAPTCHAs present a relatively small and finite set of challenge variations. Theoretically, it is possible to collect all possible variations of these CAPTCHAs, and the design of a new CAPTCHA under this type requires the generation of new elements within the CAPTCHAs (such as a new animal type in Rotation CAPTCHAs). 
Conversely, the second category consists of ``Dynamic Complexity'' CAPTCHAs, exemplified by Geetest visual reasoning CAPTCHAs. These CAPTCHAs are characterized by their generation of challenges through the combination of multiple elements or objects in various configurations. Due to this dynamic complexity, the number of potential CAPTCHA variations is vast and continually evolving, making it impractical to collect all possible challenges.

\subsection{LLMs for Solving CAPTCHAs (RQ2)}
We further evaluate if LLMs can be used to solve existing reasoning CAPTCHA problems, and if not, what are the key challenges that hinder their application. 

\subsubsection{Evaluation Strategy}
To rigorously evaluate the effectiveness of LLMs in solving reasoning CAPTCHAs, we select two state-of-the-art models, GPT-4~\cite{gpt4} and Google Gemini~\cite{gemini}. Our evaluation is designed to probe two critical facets of LLM performance: (1) the models' capacity to accurately comprehend the given CAPTCHA task and (2) their proficiency in methodically executing the necessary steps to resolve the CAPTCHA challenge. We employ two distinct approaches to test these capabilities: zero-shot prompting~\cite{liu2023pre} and Chain-of-Thoughts (CoT) strategy~\cite{wei2022chain}. These methodologies have been previously applied to diverse reasoning tasks, including symbolic~\cite{wei2023chainofthought} and mathematical reasoning~\cite{chen2023theoremqa}, offering a proven framework for assessing LLMs. 

\begin{figure*}[t]
    \centering
    \includegraphics[width=\textwidth]{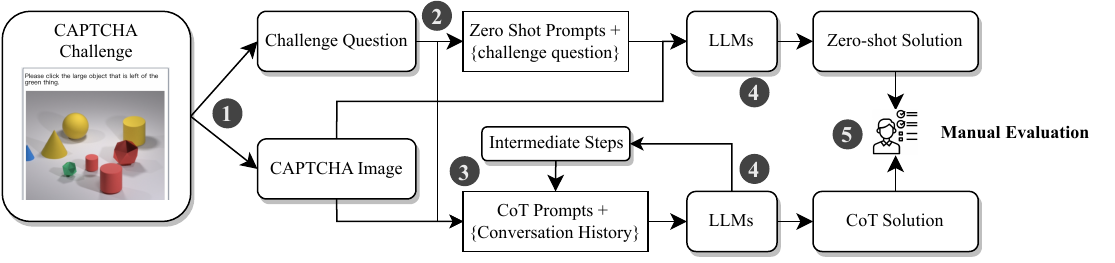}
    \caption{The proposed strategy to test LLMs on CAPTCHA solving.}
    \vspace{-10pt}
    \label{fig:testing-strategy}
\end{figure*}

Our strategy is detailed in Figure~\ref{fig:testing-strategy}. Initially, we \ding{182} prepare each CAPTCHA by isolating its image section with the question texts, employing translation to English when necessary. 
We then evaluate the LLM's capabilities in CAPTCHA solving with two strategies.
\ding{183} Zero-shot methodology~\cite{liu2023pre}: Here, the LLM is directly prompted with the CAPTCHA image and associated question. To facilitate this, the text portion of the CAPTCHA is integrated into a straightforward prompt (``Please examine the following CAPTCHA challenge and provide the step-by-step solution''), designed to elicit a direct solution from the model.
\ding{184} Chain-of-Thoughts (CoT) methodology~\cite{wei2023chainofthought}: we prompt the LLM to navigate the challenge through a series of iterative and conversational steps. This approach systematically breaks down the CAPTCHA challenge into smaller and more manageable components, each addressed in sequence to construct a comprehensive solution. Detailed guidelines for this process and specific prompt templates are provided in Appendix~\ref{sec:appendix:prompts}.
\ding{185} For both zero-shot and CoT methods, we utilize the respective multimodal API endpoints of the models (\texttt{gpt-4-vision-preview}\cite{gpt4-p} and \texttt{gemini-pro-vision}\cite{gemini-p}), submitting the CAPTCHA image and question as input.
\ding{186} Upon receiving responses from the LLM, we conduct a manual review to determine the correctness of each solution. To enhance experimental efficiency, we employ a manual correction strategy: we continue the testing if a substep in CoT is correctly executed; otherwise, we label the failed trial and provide the correct solution for the LLM to continue on the next substep. This process is repeated until the CAPTCHA is fully resolved.

This strategy offers two primary advantages: firstly, it facilitates a direct comparison between the effectiveness of zero-shot and CoT approaches in CAPTCHA solving. Secondly, the manual correction mechanism allows us to precisely evaluate the LLMs' performance on each sub-step of the CAPTCHA  solving process, ensuring that the analysis is not hindered by failures in preceding steps.

\subsubsection{Experiment Setup}

We meticulously select 10 tasks from each identified sub-category of CAPTCHA, resulting in a comprehensive test set of 50 CAPTCHAs. This selection is made to ensure a balanced representation of various CAPTCHA  types. To reduce randomness, we repeat each experiment 10 times, resulting in a total number of 1000 trials (i.e., 2 models * 5 types * 10 CAPTCHAs * 10 repetitions). 
The experiments are conducted following the previously outlined strategies, focusing on several key metrics to assess the performance of the LLMs:

\begin{enumerate}[leftmargin=*,itemsep=1pt,topsep=0pt,parsep=1pt]
    \item \textbf{Task Comprehension Correct Rate:} The primary metric assessed is the ability of LLMs to correctly comprehend the task presented by each CAPTCHA. This is determined by manually evaluating whether the LLMs exhibit an accurate understanding and appropriate intent to solve the given CAPTCHA  challenge, based on its image and description.
    \item \textbf{Success Rate Analysis:} We closely monitor and measure the success rate of LLMs in solving the diverse categories of reasoning CAPTCHAs. We manually review the responses generated by the LLMs, with a focus on assessing the correctness of their solutions. It is important to note that for the zero-shot approach, a solution is deemed correct only if the LLM response outlines the accurate procedure for resolving the CAPTCHA. In contrast, for the CoT approach, a sub-step is considered successful if the solution proposed by the LLM for that specific sub-step is accurate.
    \item \textbf{Failure Reason Analysis:} Whenever an LLM fails to solve a CAPTCHA, we meticulously analyze the root causes of the failures. This involves delving into the LLM's processing and decision-making mechanisms to pinpoint common issues, e.g., misinterpretation of the task, inadequate reasoning capabilities, or errors in logical deduction. 
\end{enumerate}




\begin{table*}[t]
\resizebox{\textwidth}{!}{
\begin{tabular}{c||cc|cc||cc|cc|cc|cc}
\hline
\multirow{2}{*}{\textbf{Approach}} & \multicolumn{4}{c||}{\textbf{Zero Shot}} & \multicolumn{8}{c}{\textbf{CoT}} \\ \cline{2-13} 
                          & \multicolumn{2}{c|}{\multirow{2}{*}{\centering Task Understanding}} & \multicolumn{2}{c||}{\multirow{2}{*}{\centering Success Rate}} & \multicolumn{2}{c|}{\multirow{2}{*}{\centering Task Understanding}} & \multicolumn{2}{c|}{\multirow{2}{*}{\centering Average \# of subtasks}} & \multicolumn{2}{p{2.8cm}|}{\centering Average \# of successful subtasks} & \multicolumn{2}{c}{\multirow{2}{*}{\centering Success Rate}} \\ \cline{1-13}

Model & GPT-4 & Gemini & GPT-4 & Gemini & GPT-4 & Gemini & GPT-4 & Gemini & GPT-4 & Gemini & GPT-4 & Gemini \\ \hline
Angular         & 37.9\% & 20.7\% & 0.0\%       & 0.0\%       & 68.1\% & 53.4\% & 4.7 & 5.1 & 0.0   & 0.1 & 0.0\%      & 0.0\%       \\
Gobang          & 93.1\% & 81.0\% & 2.6\%  & 0.0\%       & 96.5\% & 95.7\% & 5.6 & 4.3 & 2.8 & 1.3 & 4.5\%  & 3.3\%  \\
IconCrush       & 94.8\% & 88.8\% & 4.2\%  & 0.0\%       & 97.4\% & 96.6\% & 4.6 & 4.7 & 3.7 & 1.4 & 7.9\%  & 4.0\%  \\
Space           & 94.6\% & 94.8\% & 44.0\% & 37.2\% & 98.3\% & 95.7\% & 4.1 & 3.1 & 3.3 & 2.2 & 55.2\% & 50.5\% \\
Space Reasoning & 59.5\% & 53.4\% & 34.3\% & 25.9\% & 90.3\% & 89.7\% & 3.1 & 3.2 & 2.2 & 1.8 & 37.3\% & 27.8\% \\ \hline
Average         & 76.0\% & 67.8\% & 16.5\% & 12.0\% & 90.1\% & 86.2\% & 4.4 & 4.1 & 2.4 & 1.4 & 21.0\% & 17.1\% \\ \hline

\end{tabular}}
\caption{Experimental results of applying the testing strategy over the selected CAPTCHAs.} 
\label{tab:empirical-study} 
\end{table*}

\subsection{Findings}

The experimental results of our empirical study are detailed in Table~\ref{tab:empirical-study}. From these results, several intriguing findings emerge. Firstly, it is evident that LLMs are capable of understanding the task of CAPTCHA solving, as they consistently aim to generate plausible outcomes for solving CAPTCHAs. This is demonstrated by the high Task Understanding Rate, with 90.1\% for GPT-4 and 86.2\% for Gemini. Despite this understanding, both LLMs exhibit notably poor performance in actually solving reasoning CAPTCHAs. Specifically, their success rates in solving the five types of CAPTCHAs are exceedingly low, at only 16.0\% for GPT-4 and 12.0\% for Gemini. When employing the CoT approach, there is a marginal improvement in performance, reaching 21.0\% for GPT-4 and 17.1\% for Gemini, but it still remains evident that the models are unable to correctly solve reasoning CAPTCHAs autonomously without human intervention.

\begin{tcolorbox}[colback=gray!25!white, size=title,breakable,boxsep=1mm,colframe=white,before={\vskip1mm}, after={\vskip0mm}]
\textbf{Finding 1:} Despite the ability to understand the tasks within reasoning CAPTCHAs, the two state-of-the-art LLMs, GPT-4 and Gemini, are unable to effectively solve these tasks without human guidance.
\end{tcolorbox}

We further examine the CoT-solving process for both models to ascertain the reasons behind their failures. Notably, while the success rates using the CoT approach are low, both models demonstrate a significant capability in resolving a substantial portion of the generated subtasks in the CAPTCHA-solving process. For instance, GPT-4 generates an average of 4.4 subtasks across five different types of tasks and successfully solves 2.4 of them. This represents a substantial 54.5\% success rate for individual subtasks, which is significantly higher than the overall task completion rate of 21.0\%. This indicates that while LLMs can process and reason through the CAPTCHA-solving steps in a manner akin to humans, they often falter at specific stages. However, in a sequential solving process, the failure of a single subtask inevitably leads to the failure of the overall task. 

\begin{tcolorbox}[colback=gray!25!white, size=title,breakable,boxsep=1mm,colframe=white,before={\vskip1mm}, after={\vskip0mm}]
\textbf{Finding 2:} LLMs could meaningfully decompose a CAPTCHA challenge into subtasks and solve a large portion of them. However, due to the incapability of resolving some subtasks, they fail to complete the reasoning process as a whole.
\end{tcolorbox}

\begin{table}[h]
\small
\centering
\resizebox{\linewidth}{!}{
\begin{tabular}{l||c|c}
\hline
\textbf{Subtask Category}                                  & \textbf{unique subtasks} & \textbf{Success Rate} \\ \hline
Understand the task                               & 6 & 91.8\%             \\ \hline
Single-criteria object searching            & 12 &      80.8\%        \\ \hline
Multi-criteria object searching   & 6 &     25.0\%         \\ \hline
Orientation identification                 & 2 &     32.4\%         \\ \hline
Single-condition Judgement          &   8    &     78.2\%         \\ \hline
Multi-condition Judgement & 3 &      22.4\%        \\ \hline
Others                                            & 4 &     12.0\%         \\ \hline
\end{tabular}}
\caption{Success Rate of Subtasks.}
\label{tab:subtasks-rate}
\end{table}

To delve deeper into the nature of both successful and unsuccessful subtasks in the CAPTCHA-solving process, we inspect all subtasks generated during the experiment and categorize them based on their characteristics. We then count the number of unique subtasks generated in this process, and document their rate of being successfully completed by the LLMs in Table~\ref{tab:subtasks-rate}. 
It is noticed that LLMs try to complete the CAPTCHA challenge with multiple types of subtasks, such as identifying or localizing an object based on certain attributes (color, shape, etc.) and identifying its orientation. 
While LLMs exhibit the capability to complete these tasks, their performance is notably better when constrained to a single criterion in different types of tasks. In particular, they can complete single-criteria object searching with a success rate of 80.8\%, while it is difficult to handle the case with two or more criteria provided with a success rate of 25.0\%. The same conclusion holds for the case when the LLMs are prompted to judge if a declaration holds or not (78.2\% vs 22.4\%.)
An illustrative example is that in the 3D logical CAPTCHA (Figure~\ref{fig:3d2}), LLMs can proficiently identify all cylinders. However, when asked to locate a specific red cylinder, they may hallucinate and indicate an incorrect location, or even point to non-cylinder objects.

\begin{tcolorbox}[colback=gray!25!white, size=title, breakable, boxsep=1mm, colframe=white, before={\vskip1mm}, after={\vskip0mm}]
\textbf{Finding 3:} LLMs demonstrate proficiency in recognizing and understanding natural objects within CAPTCHAs, but their performance significantly diminishes with abstract objects and multi-criteria tasks, revealing a limitation in their cognitive processing capabilities.
\end{tcolorbox}

Lastly, LLMs exhibit difficulties in processing long instructions encompassing multiple steps, demonstrating an inability to recall outcomes of previous sub-steps, even when the complete conversation history is provided at the LLM endpoints. This challenge could potentially stem from the models' attention-shifting dynamics~\cite{vaswani2023attention, yang2023chatgpt} as revealed in the previous research, which seemingly prioritize the most recent conversational inputs over earlier ones. Additionally, the propensity of LLMs to generate hallucinations increases when they are presented with multiple instructions within a single prompt, complicating their task-solving effectiveness.

\begin{tcolorbox}[colback=gray!25!white, size=title, breakable, boxsep=1mm, colframe=white, before={\vskip1mm}, after={\vskip0mm}]
\textbf{Finding 4:} LLMs struggle with complex and multi-step instructions, leading to challenges in task continuity and an increased likelihood of hallucinations when handling multiple directives simultaneously.
\end{tcolorbox}

The above empirical study conclusively demonstrates that while LLMs exhibit promising capabilities in certain aspects of CAPTCHA solving, they also face significant challenges with multi-step instructions and maintaining task continuity. In light of these findings, our objective is to strategically harness the strengths of LLMs to devise a methodical approach for CAPTCHA resolution. This is achieved with a new DSL and automated framework, as detailed below.

%% file: Tex/4-Methodology.tex
\section{CAPTCHA Domain Specific Language}\label{sec:dsl}

\subsection{Motivation}
Our empirical study highlights that LLMs can potentially tackle reasoning CAPTCHAs using the CoT strategy, which decomposes the challenge into smaller subtasks and addresses them in a divide-and-conquer manner. However, the varying levels of LLM proficiency across different subtasks introduce instability into the challenge-solving process. This variability leads us to consider whether it is feasible to formalize the CAPTCHA  solving process into a series of steps, each individually tailored to be within the LLM's capabilities.

This consideration has inspired the development of our CAPTCHA  Domain Specific Language (DSL). This new DSL is designed to formally outline the operations involved in solving CAPTCHAs, essentially \textit{encoding the CAPTCHA solving process with a series of LLM-easy tasks}. The creation of CAPTCHA  DSL offers several significant advantages. 
First, the operations delineated by CAPTCHA  DSL align with the competencies of LLMs, ensuring that each step of the solution is approachable and solvable, thereby enhancing the overall success rate of solving actual CAPTCHAs.
Second, CAPTCHA DSL, being an abstract representation of the natural language process of CAPTCHA  solving, can be seamlessly translated back into natural language. This feature facilitates the integration of CAPTCHA  DSL into LLMs for reasoning and processing.
Third, CAPTCHA  DSL brings a level of formalization to the CAPTCHA  solving procedure. Solutions generated in this language can be rigorously verified for syntactical correctness, allowing for straightforward identification and rectification of errors.

In the following, we elaborate on the formal definition of CAPTCHA  DSL and showcase its application through illustrative examples. This demonstration will underscore the practical utility and effectiveness of CAPTCHA  DSL in streamlining and enhancing the process of solving reasoning CAPTCHAs with LLMs.

\subsection{High-level Structure}
We first delineate the high-level structure and components of the CAPTCHA DSL. As exemplified in Figure~\ref{fig:dsl-example}, a CAPTCHA DSL script comprises multiple lines of statements, with each line representing a specific operation targeting elements within the CAPTCHA solving context. This structure is reminiscent of the SQL~\cite{sql} syntax, where scripts are composed of clauses, the fundamental building blocks of statements. Each clause in the CAPTCHA DSL is constructed following specific syntax rules and can include a variable number of components. Below we detail the components of the CAPTCHA DSL.

\begin{enumerate}[leftmargin=*,itemsep=1pt,topsep=0pt,parsep=1pt]
    \item \textbf{\texttt{Keywords:}} The operational keywords function as the core unit operations in CAPTCHA DSL. There are four main keywords: \texttt{search}, \texttt{reason}, \texttt{locate}, and \texttt{operate}. They correspond to the atomic operations that our empirical study (Section~\ref{sec:study}) has shown to be effectively executed by LLMs, i.e. searching objects from the CAPTCHA given requirements, reasoning for a certain task, or performing operations on the actual CAPTCHA.

    \item \textbf{\texttt{Objects:}} This component represents the tangible items in the CAPTCHA image, such as animals in the rotation CAPTCHA (Figure~\ref{fig:rotation2}) and emojis in the Bingo CAPTCHA (Figure~\ref{fig:bingo2}). 

    \item \textbf{\texttt{Attributes:}} These are properties of \texttt{objects}. In CAPTCHA DSL, attributes are limited to a subset that is effective for actual CAPTCHA solving, as informed by our empirical study. This includes characteristics like \textit{size}, \textit{color}, \textit{type}, \textit{orientation}, among others.

    \item \textbf{\texttt{Predicates:}} These are conditions that can be evaluated to three-valued logic (3VL)~\cite{3vl} (true/false/unknown) or Boolean values. They are used to constrain the effects of expressions. Predicates function through  the standardized set of boolean operators. 

    \item \textbf{\texttt{Expressions:}} Expressions such as `=' in CAPTCHA DSL consist of components that operate based on specific logic to yield Boolean or attribute values. These expressions can be used for updating attributes or, in conjunction with predicates, evaluating conditions.

    \item \textbf{\texttt{Descriptions:}} A special component in CAPTCHA DSL is the natural language descriptions that can be provided as variables to the \texttt{reason} keyword, which can be arbitrarily provided and handled by the LLM. 
\end{enumerate}

\subsection{Syntax Constraints}
The above CAPTCHA DSL is regulated through a loosely defined syntax, which is primarily designed to fulfill the conclusions drew from the empirical study. Similar to Python and other programming languages, it is requested that all the variables appearing in the statements are properly defined, and their values are properly updated without use-before-definition. In addition, we define two additional syntax rules to bound the keywords to regulate that each line of the statement in the CAPTCHA DSL script can be translated into a natural language task, and the task is proved to be highly achievable by LLMs as listed in Table~\ref{tab:subtasks-rate}. (1) The operational keyword \texttt{reason} is used by LLMs to perform easy reasoning tasks given an \texttt{Description}. To ensure that the task is an `easy' reasoning task, we only allow the reasoning result to be a 3VL logic, or an \texttt{orientation} attribute of an object. This maps to the empirical study, where we realize that LLMs are better at judging if the condition is correct or wrong, but not good at making complex reasonings. (2) The keyword \texttt{search} should always and only be bounded by single attributes. This aligns to the fact that LLMs can well tell the objects from the images when only one filtering criteria is given, yet they cannot perform well when more than one is provided. 

The above DSL syntax constraints enables a local verification of the unbounded scripts. Given a DSL script, we can parse the prompts through the above DSL syntax constraints and validate if the constraints are met, and if the script fulfills the variable definition logic (such as if the value is not properly defined).

\subsection{A Running Example}\label{sec:dsl:example}

We present a detailed running example to illustrate the efficacy and modeling capabilities of our CAPTCHA DSL in the context of a 3D CAPTCHA shown in Figure~\ref{fig:3d1}. The DSL generated by GPT-4 is presented in the following Figure~\ref{fig:dsl-example}. This 3D CAPTCHA challenge requires the solver to identify and reason about the spatial orientation of alphanumeric characters in a three-dimensional space. Specifically, the task involves finding a number `0' that is oriented in the same direction as the letter `W.'. Our DSL strategy streamlines this complex problem into a series of logical steps, each represented by a single line of instruction within our DSL script, as shown below.

\begin{figure}[t]
	\centering
	\includegraphics[width=\linewidth]{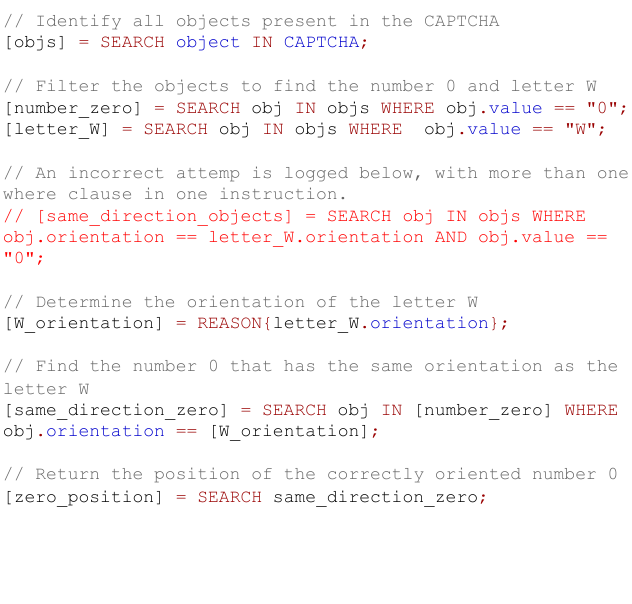}
 \caption{Example DSL script for 3D CAPTCHA challenge.}
	\label{fig:dsl-example}
\end{figure}

Initially, the script identifies all the objects within the CAPTCHA environment, creating a foundational dataset from which specific objects can be filtered and analyzed. The DSL then proceeds to filter the objects to isolate instances of the number `0' and the letter `W' each through a separate search operation as mandated by the DSL's syntax constraints. An incorrect attempt is deliberately included to showcase the script's syntax in red color: a line with multiple WHERE clauses is flagged, reflecting an implementation that does not conform to the DSL's syntactic rules. This error can be identified through the local syntax checker. 

Following the DSL's structured approach, the script deduces the orientation of the letter `W' through a reasoning operation. With this information, it then conducts a search for the number `0' that shares this orientation. Finally, the DSL script concludes by locating the correct position of `0' that aligns with 'W,' thus demonstrating the potential for the DSL to effectively model the solution to a CAPTCHA challenge. Through this example, we highlight the precision and clarity of our DSL, alongside its inherent ability to self-validate and prevent syntactically incorrect implementations that could otherwise impede the CAPTCHA solving process.

\section{Methodology}\label{sec:methodology}

\begin{figure*}[t]
    \centering
    \includegraphics[width=0.95\textwidth]{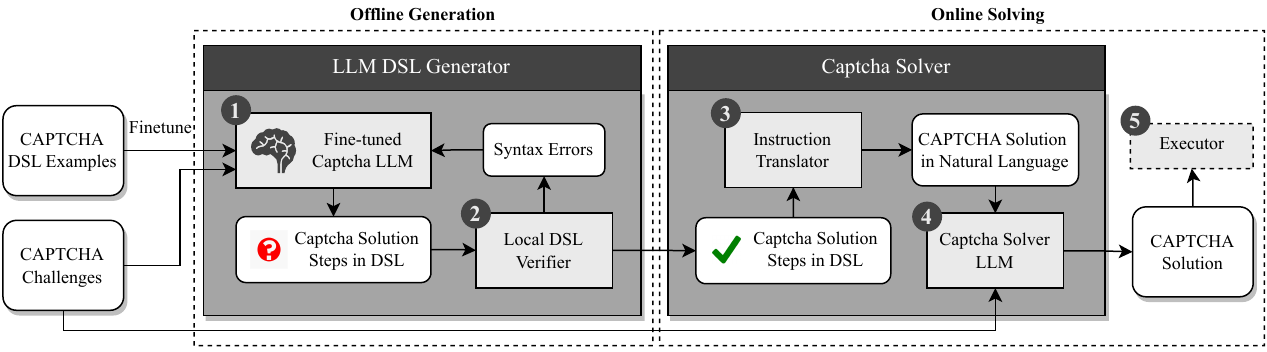}
    \caption{Overview of \tool{}. }
    \label{fig:overview}
\end{figure*}

\subsection{Overview}\label{sec:overview}

In the pursuit of an solution for solving complex and commercial-level reasoning CAPTCHAs, we introduce \tool{}\footnote{Oedipus is the renowned Greek mythological figure known for solving the riddle posed by the Sphinx.}, a comprehensive automated framework that harnesses the capabilities of multi-modal LLMs. Figure~\ref{fig:overview} provides a schematic overview of \tool{}. The cornerstone of this framework is to leverage the custom designed CATPCHA DSL illustrated in Section \ref{sec:dsl}, to systematically articulate the CAPTCHA solving process. Within this DSL, each operation is carefully aligned with the actions that LLMs have high confidence to effectively execute, a strategy grounded in the insights gained from our prior empirical study.

\tool{} operates through the following structured procedures. It contains two main phases: task generation and solution generation. They can be further detailed into the following steps. \ding{182} \textit{Pre-Solving Preparation}: In the initial phase, a custom LLM undergoes fine-tuning with DSL examples to master its syntax and structure. Faced with a new CAPTCHA challenge, this adept LLM generates a DSL script, meticulously delineating the steps necessary to unravel the CAPTCHA challenge.
\ding{183} \textit{Syntax Verification}: The crafted DSL script is subjected to rigorous verification in a local DSL verifier. This step is crucial for identifying and flagging any syntactical inaccuracies. Detected errors trigger a feedback loop to the CAPTCHA LLM, instigating another round of script generation, informed by the identified syntax errors.
\ding{184} \textit{Solution Translation}: Upon achieving a syntax-error-free DSL script, it proceeds to the instruction translator. This component transforms the DSL script into a format comprehensible in natural language, tailor-made for processing by LLMs.
\ding{185} \textit{CAPTCHA Solving}: The natural language translation of the CAPTCHA solution, coupled with the original CAPTCHA challenge, is then processed by the multimodal CAPTCHA solver LLM. This final phase yields the ultimate solution to the CAPTCHA.
\tool{} leverages the analytical prowess of LLMs within the structured confines of the DSL. This harmonization is aimed at dissecting and resolving reasoning CAPTCHAs methodically with better accuracy. In the following of this section, we detail the design of each module.

\subsection{CAPTCHA DSL Script Generation}\label{sec:method:generation}

Our investigation reveals that LLMs possess the capability to comprehend CAPTCHA challenges and generate meaningful step-by-step solutions using the CoT approach. The creation of the CAPTCHA DSL aims to guide LLMs in generating solutions that adhere to specific rules and structures, thereby enhancing the likelihood of successfully completing each step. To facilitate this, we have developed a method for instructing an LLM to automatically generate a CAPTCHA DSL program tailored to a particular type of CAPTCHA challenge.

This process involves three steps. First, we manually generate correct DSL scripts for a given CAPTCHA challenge. This step is straightforward and takes up to 5 minutes for each CAPTCHA challenge given the authors of this work is familiar with the DSL language. Second, we utilize the fine-tuning technique to develop an LLM that specializes in CAPTCHA DSL program generation. The DSL examples generated in previous step are employed as training data to fine-tune the LLM, enhancing its proficiency and accuracy in generating CAPTCHA DSL programs. Third, we employ few-shot prompting with the LLM, using the sample CAPTCHA DSL programs as illustrative examples. This process is exemplified in the textbox provided below. In practice, we find that this generation task can be effectively completed by fine-tuned code-llama~\cite{ftllama} or GPT-3.5 with a minimum number of 20 pieces of examples, or more powerful models such as GPT-4 without any fine-tuning process. In Appendix~\ref{sec:appendix:prompt-example}, we provide a concrete example on how to instruct the LLM to generate CAPTCHA DSL scripts. 

Once a DSL script is generated by the LLM, it undergoes a thorough examination in a local DSL Verifier. This verifier plays a crucial role in ensuring the syntactical correctness of the script. It checks for compliance with the DSL's syntax rules and verifies that each operation within the script is feasible and logical within the context of the specific CAPTCHA challenge. If any syntax errors or logical inconsistencies are detected, the verifier flags these issues as illutrated in Section~\ref{sec:dsl}, which are then used as feedback to refine the LLM's future script generation.

This verification step not only ensures the accuracy and reliability of the generated DSL script but also provides valuable insights for further fine-tuning the LLM to enhance its CAPTCHA solving capabilities. Through this iterative process of generation, verification, and refinement, the LLM becomes increasingly adept at creating effective and accurate CAPTCHA DSL scripts, thereby streamlining the CAPTCHA solving process.

\subsection{Instruction Translation}

Following the generation of the CAPTCHA DSL script, the subsequent phase entails translating these DSL instructions into natural language. This translation is pivotal, as it converts the structured DSL commands into a format that is understandable and actionable by the CAPTCHA Solving module, which then proceeds to address the specific CAPTCHA challenge. Given the critical role of this translation in bridging the gap between the DSL's structured syntax and the LLM's interpretative flexibility, it is paramount to ensure the accuracy and precision in this translation.

To achieve the highest level of translation fidelity, we employ an LLM, utilizing two strategies for optimal results. First, we leverage prompt engineering and meticulously craft prompts that include the definitions of our DSL syntax as described in Section~\ref{sec:dsl}, thereby orienting the LLM to accurately grasp and interpret the DSL instructions. This foundational understanding is crucial for the LLM to correctly translate the DSL script into natural language instructions. Second, similar to the multi-shot prompting strategy adopted in the DSL generation phase to furnish the LLM with examples demonstrating the translation of instructions into DSL scripts, we implement a reverse process for this phase. By providing examples that illustrate how DSL scripts can be effectively mapped back to natural language instructions, we facilitate a deeper comprehension by the LLM of the intended semantic and functional translation. Through practical application, we observe that combining prompt engineering with this reversed example provision markedly enhances the accuracy of the instruction generation process, ensuring that the CAPTCHA Solving module receives clear, precise, and actionable directives to solve the challenges at hand.

\subsection{CAPTCHA Solving}

With the natural language instructions derived from the DSL script, we proceed to the CAPTCHA solving phase. In this step, we use these instructions as CoT prompts to guide the CAPTCHA LLM through the actual CAPTCHA challenge. This process is straightforward but methodical: the instructions, along with the original challenge image, are fed into the CAPTCHA LLM. The LLM then follows these step-by-step instructions to systematically tackle and solve the CAPTCHA.

As an additional optional step, the solutions generated by the LLM can be integrated into an automated executor. This executor is programmed to interact with the CAPTCHA on the target website, inputting the solution directly and completing the challenge. This automation step, while optional, can streamline the process and eventually achieve automated CAPTCHA solving. 

\subsection{Discussion}

Our framework is intentionally not calibrated for achieving an exceptionally high success rate in CAPTCHA solving. This design philosophy aligns with the CAPTCHA evaluation standard proposed by Microsoft~\cite{microsoft-hips}, which suggests comparing the economic cost of automated CAPTCHA solving against the cost of human labor for the same task. While human solvers are expected to achieve a solving rate of above 90\%~\cite{microsoft-hips, von2003captcha}, an automated approach like ours could be deemed equally effective if it maintains a 50\% success rate, provided its operational cost is less than one-fourth of the laber cost of hiring a human solver ($1-0.5^4 = 93.75\%$).
Given this cost-benefit analysis, \tool{} is designed to optimize performance within these economic constraints, eschewing complex and cost-incurring enhancements in favor of a more streamlined approach. Consequently, \tool{} does not incorporate an active feedback mechanism to verify the correctness of its substep solutions during the reasoning process. This decision stems from the practical challenge of confirming whether the LLM has accurately identified objects without reliable external references for validation. While iterative identification and majority voting could theoretically improve accuracy, the additional computational and financial costs of increased API queries make this option untenable. Therefore, our strategy intentionally forgoes such enhancements to maintain the economic viability and operational simplicity of \tool{}.

%% file: Tex/5-Evaluation.tex
\section{Evaluation}\label{sec:evaluation}

We evaluate the performance of \tool{} on real-world reasoning CAPTCHA challenges. In particular, we are interested in four research questions. 

\begin{itemize}[leftmargin=*,itemsep=1pt,topsep=0pt,parsep=1pt]
\item\textbf{RQ1 (Effectiveness)} How effective of \tool{} in addressing real-world reasoning CAPTCHAs automatically?

\item\textbf{RQ2 (Ablation)} How effective is each strategy contributing to the success of \tool{}?

\item\textbf{RQ3 (Transferability)} Can \tool{} resolve new CAPTCHA tasks that are not included in our empirical study, i.e. its transferability to new CAPTCHAs?
\end{itemize}


\subsection{Experimental Setup}

\noindent\textbf{Evaluation Baselines.} 
We implement \tool{} with 1,554 lines of Python3 codes. 
In our assessment of \tool{} under varied conditions, we incorporate three multimodal LLMs that are currently accessible: OpenAI GPT-4, Google Gemini, and miniGPT-4~\cite{minigpt4}. The inclusion of GPT-4 and Gemini aligns with our preceding empirical study to ensure consistency in model evaluation, while miniGPT-4 is chosen for its prominence as the most widely recognized open-source multimodal LLM to date. For all three models, we adjust the LLM response temperature to 0, aiming to minimize the output variability and ensure deterministic responses. 
To benchmark \tool{}'s performance, we draw a comparison with the only discussed solution to reasoning CAPTCHAs in academia, \textsc{VTT}~\cite{gao2021captcha}.
Given the absence of open-source access to \textsc{VTT} and the discontinuation of the CAPTCHA challenges it was tested on, we endeavor to reconstruct their approach as accurately as possible to facilitate a fair comparison. 

\noindent\textbf{Evaluation Datasets.} In the absence of existing open-source datasets for reasoning CAPTCHAs, we embark on a systematic collection process. We target three prominent security companies known for providing reasoning CAPTCHAs: Arkose Labs~\cite{arkose}, Geetest~\cite{geetest}, and NetEase Yidun~\cite{netease}. 
By collecting all types of reasoning CAPTCHAs from them designed by the end of 2022 and available online as paid commercialized API services, we finally formulate a dataset, which includes four types of CAPTCHAs covering the three categories summarized in Section~\ref{sec:study}: \textit{Arkose-Angular}, \textit{Geetest-Gobang}, \textit{Geetest-Space}, and \textit{Yidun-Space-Reasoning}\footnote{The names are given by the vendors.}. Their examples are provided in Figure~\ref{fig:evaluation-data} of Appendix.
We meticulously collect examples through their paid CAPTCHA API services, ultimately curating a dataset comprising 50 samples for each type of challenges. Note that \textit{Arkose-FunCAPTCHA} samples exhibit repetition, differing only in their starting positions, due to the limited number of combinations inherent in Rotation-type CAPTCHAs.
Following the collection phase, we manually solve all the CAPTCHA challenges to ensure each task is paired with a correct standard answer.
Since \textsc{VTT} requires additional data for training, we collect 50 additional samples with manual labels to reconstruct their models.

\noindent\textbf{Experiment Settings} For each CAPTCHA challenge, we run three versions of \tool{} and \textsc{VTT} and examine the solutions. To reduce randomness, we repeat each trial for 10 times. Thus, we conduct a total number of 8,000, i.e., 4 projects * 50 samples * 4 settings * 10 repetitions, of experiments, with a total LLM API cost of 320.55 USD\footnote{For OpenAI and Google APIs; mini-GPT4 is implemented locally on a PC with Nvidia RTX 4090.}. 

\subsection{(RQ1) CAPTCHA Solving Performance}

Our initial exploration into the effectiveness of different models in the reasoning CAPTCHA solving process reveals insightful outcomes, as summarized in Table~\ref{tab:performance}. Notably, \tool{}, when augmented with GPT-4, outperforms its counterparts across all four types of reasoning CAPTCHAs, recording an average success rate of 63.5\%. This performance is significantly superior to that of \tool{} configured with other models. Specifically, \tool{} enhanced by Gemini achieves a commendable average success rate of 51.2\%, while the version powered by miniGPT-4 trails with a success rate of 14.3\%. In comparison, \textsc{VTT} demonstrates proficiency in the space reasoning challenges, securing success rates of 55.2\% and 44.0\%, respectively, which places it slightly behind the Gemini-powered \tool{} iteration.

A notable observation from the experiments is the uniformly low success rates encountered with Angular CAPTCHAs across all strategies. A closer inspection of the LLM outputs suggests a difficulty in formally recognizing the orientation of objects, leading to selections that approximate but do not exactly match the correct orientation. Despite this challenge, \tool{} showcases commendable performance across the board, indicating that the integration of LLMs with superior reasoning abilities substantially enhances CAPTCHA solving success rates.
This underscores the pivotal role of advanced reasoning capabilities inherent in LLMs in improving automated CAPTCHA solving outcomes. The different success rates observed across various models and CAPTCHA types illuminate the critical interplay between model selection and task-specific performance, highlighting \tool{}'s potential as a versatile and effective tool for navigating the complexities of modern CAPTCHA challenges.

\begin{table}[h]
  \centering
  \resizebox{\linewidth}{!}{
  \begin{tabular}{c||cccc}
    \hline
    \textbf{} & \textbf{GPT-4} & \textbf{Gemini} & \textbf{miniGPT-4} & \textbf{VTT}\\
    \hline
    Angular Old     & 37.4\% & 31.4\% & 8.0\% & \ding{55} \\
    Gobang          & 80.2\% & 55.8\% & 16.2\% & \ding{55} \\
    Space           & 70.8\% & 60.0\% & 18.4\% & 55.2\% \\
    Space Reasoning & 65.4\% & 57.6\% & 14.6\% &  44.0\%\\
    \hline
    Average         & 63.5\% & 51.2\% & 14.3\% & 24.8\% \\
    \hline
  \end{tabular}}
  \caption{The success rate of CAPTCHA solving.}
  \label{tab:performance}
\end{table}

We proceed to evaluate the practicality of \tool{} by comparing the average cost required to successfully solve 100 reasoning CAPTCHAs, accounting for both successful and unsuccessful attempts. This analysis is juxtaposed against the quotations from two anonymous online CAPTCHA solving services, with specifics withheld for security and ethical considerations. 
The cost comparison results, detailed in Table~\ref{tab:cost}, reveal that although \tool{} supported by GPT-4 exhibits a higher success rate, its cost per 100 CAPTCHA solutions (13.1 USD) significantly surpasses that of \tool{} powered by Gemini (3.1 USD). This discrepancy is primarily due to the GPT-4 API's pricing, which is approximately ten times higher than that of the Gemini Pro API~\cite{gemini-p}. The rates offered by the two CAPTCHA solving service providers are marginally lower than those associated with \tool{}, yet they reside within the same cost magnitude.
Although \tool{} currently incurs a higher cost compared to existing CAPTCHA solving services, which utilize undisclosed techniques, there is potential for cost reduction. Moreover, people can always run \tool{} supported by local LLMs to reduce costs. 


\begin{table}[h]
  \centering
  \resizebox{\linewidth}{!}{
  \begin{tabular}{ccccc}
    \hline
    \textbf{Cost(USD)} & \textbf{Angular} & \textbf{Gobang} &   \textbf{Space} &\textbf{Space Reasoning}\\
    \hline
    \tool{}-GPT4 & 21.6 & 12.8 & 8.7 & 9.3   \\
    \tool{}-Gemini & 5.3 & 3.2 & 1.8 & 2.1  \\
    Provider-A & 5.0 & 2.0 & 0.5 & 0.5  \\
    Provider-B & 3.0 & 2.0 & 0.3 & 0.5  \\
    \hline
  \end{tabular}}
  \caption{Average cost of different tools for successfully solving 100 CAPTCHAs.}
  \label{tab:cost}
\end{table}

\subsection{(RQ2) Ablation Study}
We further investigate if the design of each process in \tool{} (i.e., DSL generation, multi-shot prompting CAPTCHA solving) can successfully improve the performance of the CAPTCHA solving process. 

Our initial investigation focuses on the impact of local DSL generation feedback on task completion efficacy. For this purpose, we establish a comparison group in which CAPTCHA DSL scripts generated during the first attempt are directly utilized for CAPTCHA solving. This approach includes a manual review to ascertain the accuracy of the DSL script generation for each task. Results of this examination are detailed in Table~\ref{tab:dsl-rate}. It is observed that the initial success rate for generating DSL scripts across the four tasks averages at 59.8\%, whereas incorporating feedback significantly enhances this rate to 93.7\%. Thus, integrating feedback into the CAPTCHA solving process markedly improves the overall success rate, underscoring the value of iterative refinement in generating effective DSL scripts for CAPTCHA resolution.

\begin{table}[h]
  \centering
  \resizebox{\linewidth}{!}{
  \begin{tabular}{ccccc|c}
    \hline
    \textbf{Success Rate} & \textbf{Angular} & \textbf{Gobang} &   \textbf{Space 1} &\textbf{Space 2} & Average\\
    \hline
    First Trial & 61.8\% & 63.4\% & 56.8\% & 57.2\% & 59.8\%\\
    Solving  & 16.2\% & 55.3\% & 47.9\% & 38.6\%  & 39.5\%\\ \midrule
    Feedback  & 91.2\% & 94.4\% & 93.2\% & 95.8\%  & 93.7\%\\ 
    Solving  & 37.4\% & 80.2\% & 70.8\% & 65.4\%  & 63.5\%\\ 
    \hline
  \end{tabular}}
  \caption{The success rate of DSL generation. }
  \label{tab:dsl-rate}
\end{table}

We then examine if task breakdown contributes to the performance of the solving process. As shown in Table~\ref{tab:subtask}, In this table, gray rows highlight outcomes where LLMs attempt CAPTCHA solutions without segmenting the tasks into smaller, more manageable steps. MiniGPT-4 remains incapable of solving CAPTCHA problems due to its inherent limitations. In contrast, task breakdown is evidently of great assistance for GPT-4 and Gemini. This underscores the effectiveness and significance of \tool{}. In \textit{Angular} tasks, task breakdown makes it possible for LLMs to carry on such challenges, which they cannot solve directly, albeit with a modest success rate. In \textit{Gobang} and \textit{IconCrush} tasks, there is a sudden increase in success rates, elevating them from initially low levels to relatively high levels. For Space and Space Reasoning tasks, which LLMs can solve with certain levels of success rates, improvements are also observed.

\begin{table}[h]
  \centering
  \begin{tabular}{cccc}
    \hline
      & \textbf{GPT-4} & \textbf{Gemini} & \textbf{miniGPT-4} \\ \hline
    \multirow{2}{*}{Angular} & \graycell 0.0\% & \graycell 0.0\% & \graycell 0.0\% \\
                             & 37.4\%     & 31.4\%     & 0.4\%      \\ \hline
    \multirow{2}{*}{Gobang} & \graycell 2.6\% & \graycell 0.0\% & \graycell 0.0\% \\
                            & 80.2\%          & 55.7\%     & 1.6\% \\ \hline
    \multirow{2}{*}{Space} & \graycell 54.0\% & \graycell 49.2\% & \graycell 0.2\% \\
                           & 70.9\% & 60.1\% & 1.8\%  \\ \hline
    \multirow{2}{*}{Space Reasoning} & \graycell 34.3\% & \graycell 25.9\% & \graycell 0.2\% \\
                                     & 65.4\% & 57.7\% & 1.4\%  \\
    \hline
  \end{tabular}
  \caption{The impact of task breakdown. Gray row denotes no task breakdown.}
  \label{tab:subtask}
\end{table}

\subsection{(RQ3) Transferability}

\begin{figure}[htbp]
  \centering

  \subfigcapskip=3pt
  \subfigure[Arkose-FunCAPTCHA]{\adjustimage{width=0.48\linewidth, valign=c}{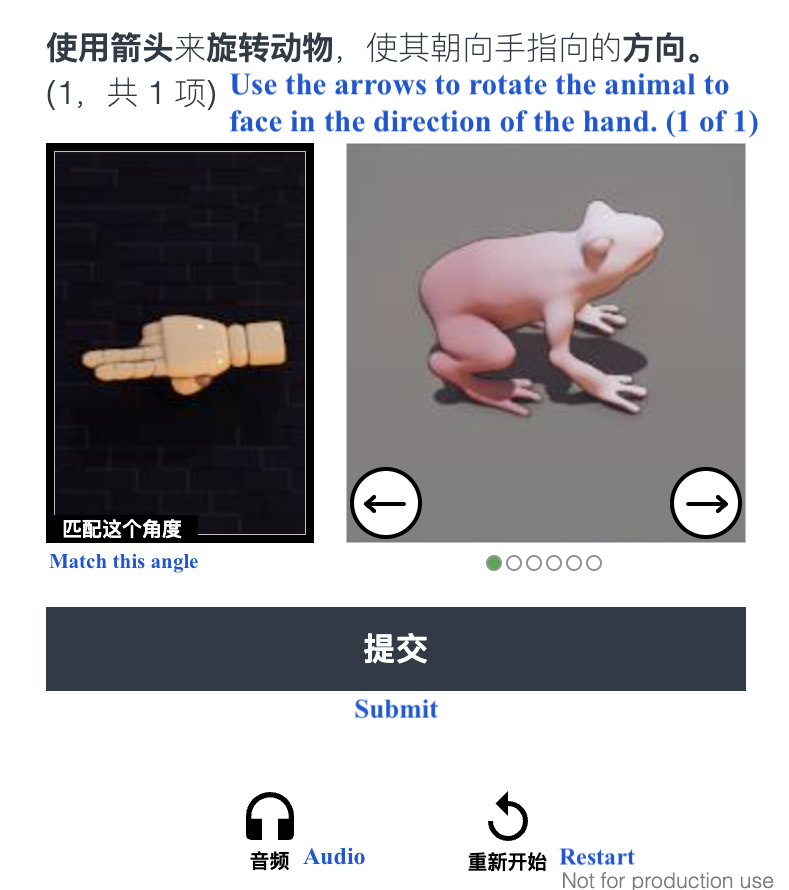}}
  \subfigcapskip=1pt
  \hfill
  \subfigure[Geetest-IconCrush]{\adjustimage{width=0.48\linewidth, valign=c}{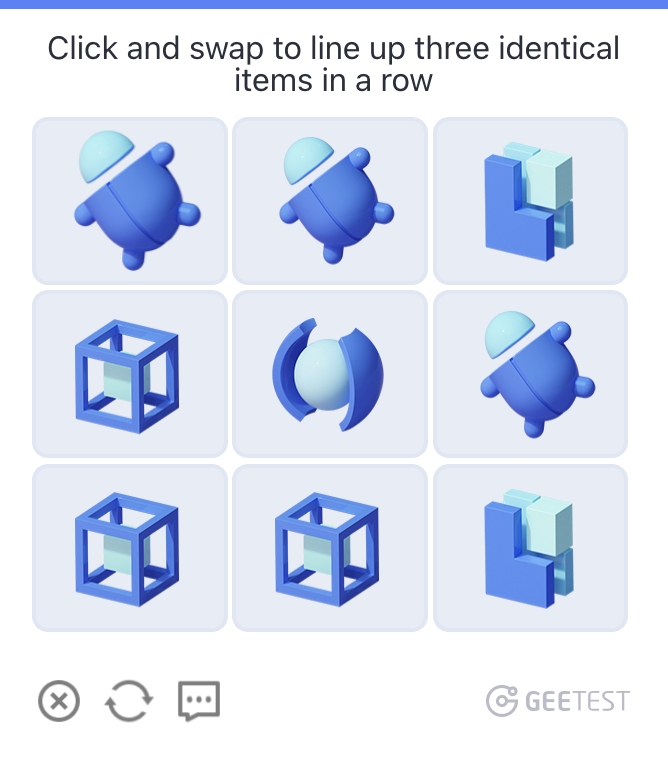}}

  \caption{Two new CAPTCHAs developed in 2023.}
  \label{fig:new-captcha}
\end{figure}

To evaluate \tool{}'s ability to adapt to emerging CAPTCHA challenges, we delve into its performance against newly designed reasoning CAPTCHAs. Following the methodology outlined in Section~\ref{sec:methodology}, which emphasizes the use of sample DSL scripts and CAPTCHA solutions for few-shot prompting, we assess \tool{}'s efficacy in navigating uncharted CAPTCHA terrains. Specifically, we focus on two novel CAPTCHA types introduced in 2023: \textit{Arkose-AngularV2} and \textit{Geetest-IconCrush}, with visual samples depicted in Figure~\ref{fig:new-captcha}. Consistent with our established experimental protocol, we curate a dataset comprising 50 instances of each CAPTCHA variant, subsequently deploying \tool{} powered by three distinct LLMs to tackle these challenges.

\begin{table}[h]
  \centering
  \resizebox{\linewidth}{!}{
  \begin{tabular}{c||cccc}
    \hline
    \textbf{} & \textbf{GPT-4} & \textbf{Gemini} & \textbf{miniGPT-4} & \textbf{VTT}\\
    \hline
    Angular New     & 35.2\% & 31.4\% & 4.2\% & \ding{55} \\
    IconCrush       & 67.4\% & 42.2\% & 6.0\% & \ding{55} \\
    \hline
    Average         & 51.3\% & 36.8\% & 5.1\% & \ding{55} \\
    \hline
  \end{tabular}}
  \caption{Transferability study on new CAPTCHAs.}
  \label{tab:performance-2}
\end{table}

The experimental outcomes, tabulated in Table~\ref{tab:performance-2}, reveal a nuanced picture of \tool{}'s adaptability. GPT-4 and Gemini, serving as the computational backbones for \tool{}, facilitate the success rates to 51.3\% and 36.8\%, respectively. In contrast, miniGPT-4 lags significantly, managing only a 5.1\% success rate. These results collectively affirm the potential of \tool{}'s pipeline to extend its applicability to CAPTCHAs beyond its initial training set, underscoring the pivotal role of LLMs' reasoning capabilities in this context. The diminished performance observed with miniGPT-4 highlights a crucial insight: the efficacy of \tool{} in addressing new CAPTCHA formats is markedly constrained by the underlying model's reasoning proficiency.

Despite the challenges encountered with miniGPT-4, the overall findings validate \tool{}'s transferability capability, illustrating its potential to remain an effective tool in the evolving landscape of CAPTCHA security. This adaptability is contingent upon continuous enhancements in LLM reasoning capabilities, ensuring that \tool{} can keep pace with the dynamic nature of CAPTCHA design and deployment.


%% file: Tex/6-Discussion.tex
\section{Discussion}\label{sec:discussion}

\subsection{Limitation}
A significant constraint of our proposed strategy is the time required for solving CAPTCHAs. We present the average solving time of \tool{} across the four benchmark reasoning CAPTCHA challenges in Table~\ref{tab:time}. Note that the solving time for \tool{}-miniGPT4 is not included because it is largely dependent on the GPU used. Notabily, the time required by \tool{}is markedly higher than that of industrial solutions currently available. Specifically, \tool{}'s processing time averages over 100 seconds for each task, a stark contrast to the sub-20-second duration required by the anonymous CAPTCHA service providers from whom we have obtained the quotations. It is worth noting that Augular and Gobang takes more time to be solved compared to other tasks.

This discrepancy highlights the efficiency gap between \tool{} and established CAPTCHA solving services, underlining the necessity for enhancements in model inference speed. Optimizations aimed at accelerating model performance could significantly reduce solving times, thereby improving \tool{}'s viability as a competitive solution in the CAPTCHA solving landscape. We are optimistic that advancements in computational efficiency and algorithmic refinements will pave the way for faster inference, making \tool{} not only a methodologically innovative but also time-efficient for CAPTCHA resolution in the future.

\begin{table}[h]
  \centering
  \resizebox{\linewidth}{!}{
  \begin{tabular}{c||cc|cc}
    \hline
    \textbf{Time(s)} & \textbf{GPT-4} & \textbf{Gemini}  & \textbf{Provider-A} & \textbf{Provider-B}\\
    \hline
    Angular         & 382.7 & 337.2 & 28.0 & 15.0 \\
    Gobang          & 124.7 & 102.2 & 20.0 & 10.0 \\
    Space           & 39.2 & 37.1 & 8.0 & 3.0 \\
    Space Reasoning & 39.3 & 37.7 & 8.0 & 3.0  \\
    \hline
    Average         & 146.5 & 128.6 & 16.0 & 7.8 \\
    \hline
  \end{tabular}}
  \caption{Time consumption of CAPTCHA solving process.}
  \label{tab:time}
\end{table}


\subsection{New CAPTCHA Design as a Defense}

The advent of LLMs has ushered in a new era where traditional CAPTCHAs, once formidable barriers to automated bots, are now increasingly vulnerable. This challenge is compounded by the realization that CAPTCHAs based on commonsense reasoning~\cite{gao2021captcha} — a concept once believed to be a strong defense—are now within the realm of problems solvable by advanced LLMs because of their broad knowledge base~\cite{zhao2023large}. In response to these developments, our proposal for enhancing CAPTCHA security involves the conceptualization of new reasoning CAPTCHA challenges from three innovative perspectives, aimed at exploiting the current limitations of LLMs.

\noindent\textbf{Complex Reasoning Chains}: The first approach involves crafting CAPTCHAs that necessitate extended reasoning chains far beyond the current processing capabilities of LLMs. For example, a CAPTCHA could present a narrative puzzle that requires understanding a multi-step logical sequence or piecing together information from various parts of a text to arrive at a conclusion. We have primarily tested that by updating the Bingo challenge in Section~\ref{tab:empirical-study} into moving two pieces, which is unsovable for current LLMs. This type of CAPTCHA can push the boundaries of LLMs' reasoning depth, requiring not just understanding individual components but synthesizing complex relationships over several logical steps. 

\noindent\textbf{Deceptive Object Recognition}: The second strategy exploits the object recognition capabilities of LLMs by introducing adversarial examples~\cite{goodfellow2014explaining} into CAPTCHA designs. For instance, CAPTCHAs could incorporate visually distorted objects that appear normal to human observers but are deliberately crafted to mislead LLM's pattern recognition. This leverages the susceptibility of neural networks to misinterpretation when faced with carefully manipulated input data.

\noindent\textbf{Unit Operations Beyond LLM Capabilities}: The third approach seeks to develop CAPTCHAs comprising unit operations that are inherently beyond the capabilities of existing LLMs. For instance, LLMs can be designed to require intuitive understanding of physical interactions in the real world, such as predicting the outcome of a complex physical event depicted in a video clip, challenging LLMs' ability to infer physical laws and dynamics from visual data alone.

Incorporating these strategies into the design of reasoning CAPTCHAs aims to elevate their security efficacy against automated solvers. However, it is imperative to acknowledge that as LLMs continue to advance in sophistication, even these novel CAPTCHA designs may eventually be surmounted. Complex reasoning chains, deceptive object recognition tactics, and challenges that currently surpass LLM capabilities could, over time, fall within the scope of what future LLM iterations can decode. This eventuality underscores the cat-and-mouse nature of CAPTCHA development and AI evolution—a continuous cycle of action and reaction, where each advancement in CAPTCHA design prompts a corresponding leap in AI problem-solving abilities.

%% file: Tex/7-Conclusion.tex
\section{Conclusion}\label{sec:conclusion}

In this work, we study the hard AI problems underlying current reasoning CAPTCHAs, and explore the automated methodology to solve these challenges through LLMs. With a custom designed CAPTCHA DSL, we design an end-to-end framework \tool{} that automatically solves reasoning CAPTCHAs. Our solution achieves an average success rate of 63.5\%, with cost comparative to commercialized CAPTCHA solving services. To propose defense, we further propose three strategies of designing more secure reasoning CAPTCHAs in the future. 

%% file: Tex/Appendix.tex
\clearpage
\appendix

\section{Appendix}\label{sec:appendix}

\subsection{Prompts for testing LLMs}\label{sec:appendix:prompts}

Below we detail the prompts for instructing the LLM to deconstruct the CAPTCHA challenge into subtasks in the CoT testing process.

\begin{tcolorbox}[colback=gray!25!white, size=title,breakable,boxsep=1mm,colframe=white,before={\vskip1mm}, after={\vskip0mm}]
\sffamily The provided image is a CAPTCHA challenge, and your goal is to \{place\_holder\}. Based on the goal, output the step-by-step solution of this type of CAPTCHA. Note that do not generate additional contents other than the instructions given above.
\end{tcolorbox}

\subsection{CAPTCHA DSL Running Example}
Below we demonstrate how CAPTCHA DSL can model a more complex CAPTCHA solving process. The following DSL script models the solution for Bingo challenge shown in Figure~\ref{fig:bingo1}.

The core logic unfolds in a loop that processes each color identified. (1) For each color, the script searches the CAPTCHA to find all objects of that specific color and locates their positions on the grid. (2) It then uses statistical reasoning to determine the most common \textit{x} (horizontal) and \textit{y} (vertical) coordinates among these objects, which helps to hypothesize about a potential line formation. (3) The decision-making process involves evaluating whether a majority of the same-colored objects share a common x or y axis, which would suggest a line either in a row or a column. If a majority is aligned along one axis, the script looks for any outlier—a piece that breaks the potential line. This outlier is identified by its unique position that differs from the common coordinate. (4) Once the outlier is found, the final step is to move this object to a position that aligns it with the other objects on the common axis, thus completing the Bingo line. 

\lstset{basicstyle=\ttfamily\small,    
breaklines=true}

\begin{lstlisting}
// Identify all objects present in the CAPTCHA
[objs] = SEARCH object IN CAPTCHA;

// Determine the colors of all objects
[colors] = REASON{[obj.color for obj in objs]};

// Iterate over each color to locate objects sharing that color
FOR c IN colors:
    candidate_objs = SEARCH obj IN CAPTCHA WHERE obj.color == c
    [candidate_objs.x, candidate_objs.y] = LOCATE candidate_objs;
    
    // Assess if a simple majority of objects share a common axis, indicating potential alignment
    common_x = REASON{MODE([obj.x for obj in candidate_objs])};
    common_y = REASON{MODE([obj.y for obj in candidate_objs])};

    // Perform reasoning to determine if the objects of the same color form a row or column
    IF REASON{COUNT(candidate_objs WHERE obj.x == common_x) > 1} AND REASON{COUNT(candidate_objs WHERE obj.y == common_y) <= 1}:
        // Majority of objects are aligned along the x-axis, find the outlier on the y-axis
        outlier = SEARCH obj IN candidate_objs WHERE obj.y != common_y;
    ELIF REASON{COUNT(candidate_objs WHERE obj.y == common_y) > 1} AND REASON{COUNT(candidate_objs WHERE obj.x == common_x) <= 1}:
        // Majority of objects are aligned along the y-axis, find the outlier on the x-axis
        outlier = SEARCH obj IN candidate_objs WHERE obj.x != common_x;
    
    // Move the identified outlier to align with other objects of the same color
    MOVE outlier TO ALIGN WITH candidate_objs ON common_axis; 
\end{lstlisting}

\subsection{DSL Script Generation Prompts}\label{sec:appendix:prompt-example}

Below we provide the prompts to guide the LLMs to generate the CAPTCHA DSL scripts, where the place holders are replaced by the actual contents as introduced in Section~\ref{sec:methodology}.

\begin{mybox}
{\textbf{\textit{DSL Generation Template}}}

Your task is to describe the process of solving a CAPTCHA in a custom domain specific language (DSL) named CAPTCHA DSL. \\

The definition of the CAPTCHA DSL is as below: \{place\_holder\} \\

We provide two examples of CAPTCHA DSLs: \{place\_holder\} below. \\

Now given the CAPTCHA task provided, please generate the soltuion with DSLs, including in-line documentation for each line of code.

\end{mybox}

\subsection{Evaluation Dataset}\label{sec:appendix:evaluation-dataset}
Below we provide four example CAPTCHA challenge as demonstrated.

\begin{figure}[htbp]
  \centering

  \subfigure[Geetest-Space]{\adjustimage{width=0.4\linewidth, valign=c}{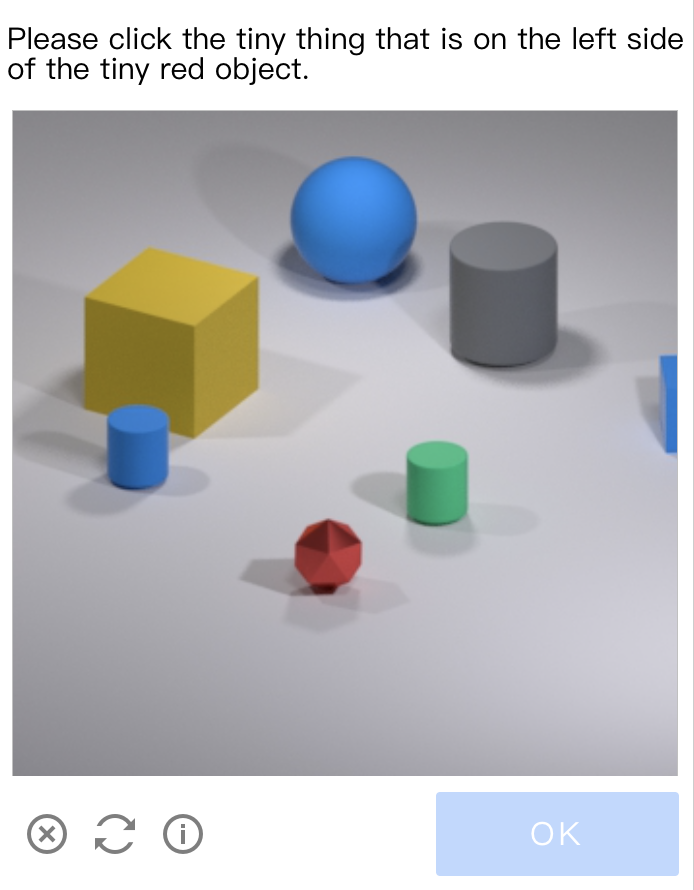}}\hspace{0.5cm}
  \subfigcapskip=7pt
  \subfigure[Geetest-Gobang]{\adjustimage{width=0.4\linewidth, valign=c}{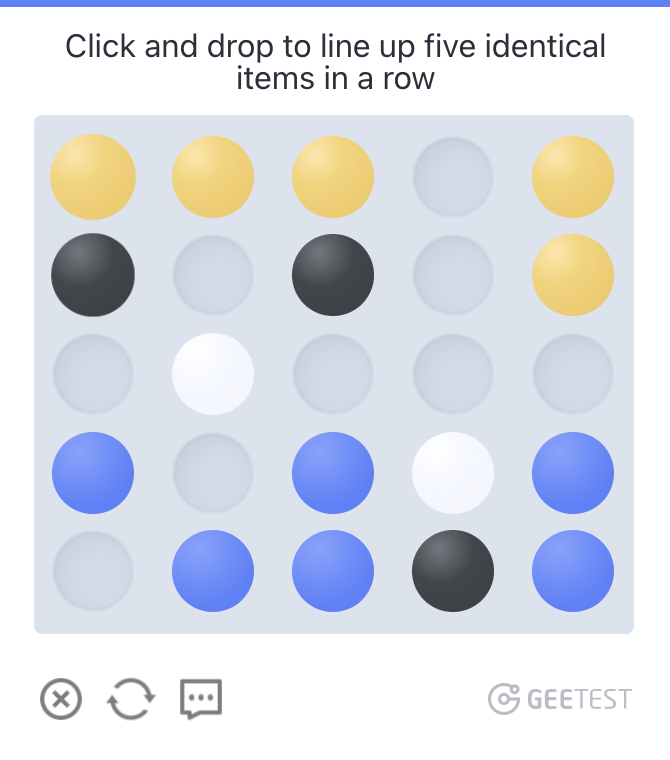}}

  \subfigcapskip=27pt
  \subfigure[Yidun-Space-Reasoning]{\adjustimage{width=0.4\linewidth, valign=c}{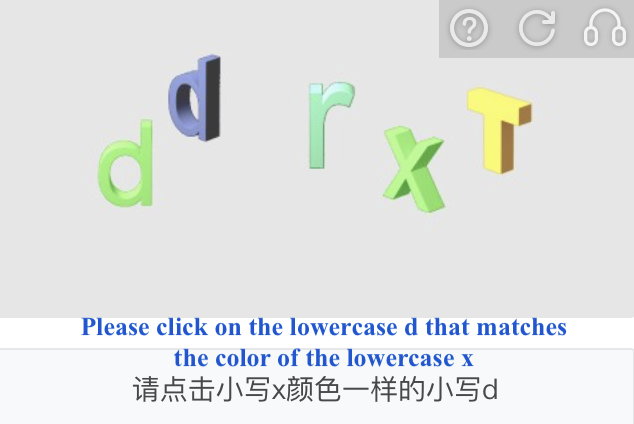}}\hspace{0.5cm}
  \subfigcapskip=5pt
  \subfigure[Arkose-FunCAPTCHA]{\adjustimage{width=0.4\linewidth, valign=c}{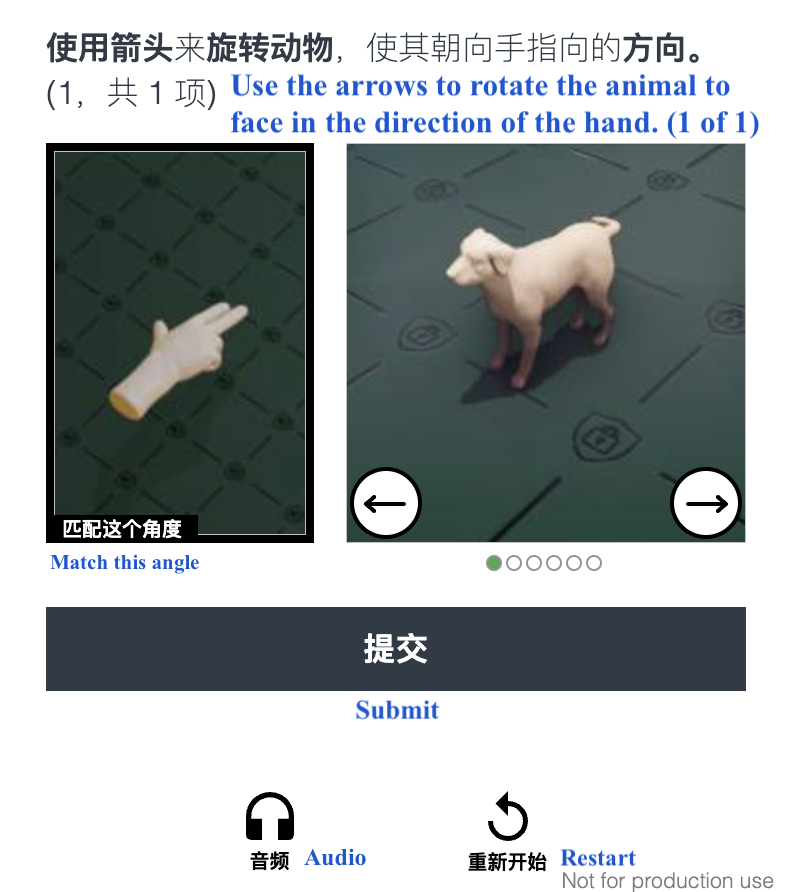}}

  \caption{Effectiveness study.}
  \label{fig:evaluation-data}
\end{figure}

\begin{figure}[htbp]
  \centering
  
  \subfigcapskip=2pt
  \subfigure[Arkose-FunCAPTCHA]{\adjustimage{width=0.4\linewidth, valign=c}{Figures/Arkose-FunCAPTCHA_new.png}}\hspace{0.5cm}
  \subfigcapskip=1pt
  \subfigure[Geetest-IconCrush]{\adjustimage{width=0.4\linewidth, valign=c}{Figures/Geetest-IconCrush.png}}

  \caption{Transferability study.}
  \label{fig:transferability-study}
\end{figure}

%% file: main.bbl
\begin{thebibliography}{10}
\providecommand{\url}[1]{#1}
\csname url@samestyle\endcsname
\providecommand{\newblock}{\relax}
\providecommand{\bibinfo}[2]{#2}
\providecommand{\BIBentrySTDinterwordspacing}{\spaceskip=0pt\relax}
\providecommand{\BIBentryALTinterwordstretchfactor}{4}
\providecommand{\BIBentryALTinterwordspacing}{\spaceskip=\fontdimen2\font plus
\BIBentryALTinterwordstretchfactor\fontdimen3\font minus \fontdimen4\font\relax}
\providecommand{\BIBforeignlanguage}[2]{{%
\expandafter\ifx\csname l@#1\endcsname\relax
\typeout{** WARNING: IEEEtran.bst: No hyphenation pattern has been}%
\typeout{** loaded for the language `#1'. Using the pattern for}%
\typeout{** the default language instead.}%
\else
\language=\csname l@#1\endcsname
\fi
#2}}
\providecommand{\BIBdecl}{\relax}
\BIBdecl

\bibitem{cap}
CAPTCHA, \url{https://en.wikipedia.org/wiki/CAPTCHA}.

\bibitem{moravec}
Moravec's\_paradox, \url{ https://en.wikipedia.org/wiki/Moravec%27s_paradox}.

\bibitem{ReCAPTCHA}
ReCAPTCHA, \url{https://www.google.com/recaptcha/about/}.

\bibitem{hCaptcha}
hCaptcha, \url{https://www.hcaptcha.com/}.

\bibitem{ye2018yet}
G.~Ye, Z.~Tang, D.~Fang, Z.~Zhu, Y.~Feng, P.~Xu, X.~Chen, and Z.~Wang, ``Yet another text captcha solver: A generative adversarial network based approach,'' in \emph{Proceedings of the 2018 ACM SIGSAC conference on computer and communications security}, 2018, pp. 332--348.

\bibitem{bursztein2014end}
E.~Bursztein, J.~Aigrain, A.~Moscicki, and J.~C. Mitchell, ``The end is nigh: Generic solving of text-based $\{$CAPTCHAs$\}$,'' in \emph{8th USENIX Workshop on Offensive Technologies (WOOT 14)}, 2014.

\bibitem{jin2023secure}
R.~Jin, L.~Huang, J.~Duan, W.~Zhao, Y.~Liao, and P.~Zhou, ``How secure is your website? a comprehensive investigation on captcha providers and solving services,'' 2023.

\bibitem{hossen2021object}
M.~I. Hossen, Y.~Tu, M.~F. Rabby, M.~N. Islam, H.~Cao, and X.~Hei, ``An object detection based solver for google's image recaptcha v2,'' 2021.

\bibitem{wang2018captcha}
H.~Wang, F.~Zheng, Z.~Chen, Y.~Lu, J.~Gao, and R.~Wei, ``A captcha design based on visual reasoning,'' in \emph{2018 IEEE International Conference on Acoustics, Speech and Signal Processing (ICASSP)}.\hskip 1em plus 0.5em minus 0.4em\relax IEEE, 2018, pp. 1967--1971.

\bibitem{gao2021research}
Y.~Gao, H.~Gao, S.~Luo, Y.~Zi, S.~Zhang, W.~Mao, P.~Wang, Y.~Shen, and J.~Yan, ``Research on the security of visual reasoning $\{$CAPTCHA$\}$,'' in \emph{30th USENIX security symposium (USENIX security 21)}, 2021, pp. 3291--3308.

\bibitem{wang2023extended}
P.~Wang, H.~Gao, C.~Xiao, X.~Guo, Y.~Gao, and Y.~Zi, ``Extended research on the security of visual reasoning captcha,'' \emph{IEEE Transactions on Dependable and Secure Computing}, 2023.

\bibitem{zong2022detrs}
Z.~Zong, G.~Song, and Y.~Liu, ``Detrs with collaborative hybrid assignments training. arxiv 2022,'' \emph{arXiv preprint arXiv:2211.12860}, 2022.

\bibitem{linkedin}
LinkedIn, \url{https://www.linkedin.com/}.

\bibitem{tiktok}
TikTok, \url{https://www.tiktok.com/}.

\bibitem{twitter}
Twitter, \url{https://www.twitter.com/}.

\bibitem{zhao2023survey}
W.~X. Zhao, K.~Zhou, J.~Li, T.~Tang, X.~Wang, Y.~Hou, Y.~Min, B.~Zhang, J.~Zhang, Z.~Dong \emph{et~al.}, ``A survey of large language models,'' \emph{arXiv preprint arXiv:2303.18223}, 2023.

\bibitem{gpt4}
GPT-4V, \url{https://openai.com/research/gpt-4v-system-card}.

\bibitem{gemini}
Gemini, \url{https://deepmind.google/technologies/gemini/#introduction}.

\bibitem{liu2023pre}
P.~Liu, W.~Yuan, J.~Fu, Z.~Jiang, H.~Hayashi, and G.~Neubig, ``Pre-train, prompt, and predict: A systematic survey of prompting methods in natural language processing,'' \emph{ACM Computing Surveys}, vol.~55, no.~9, pp. 1--35, 2023.

\bibitem{wei2022chain}
J.~Wei, X.~Wang, D.~Schuurmans, M.~Bosma, F.~Xia, E.~Chi, Q.~V. Le, D.~Zhou \emph{et~al.}, ``Chain-of-thought prompting elicits reasoning in large language models,'' \emph{Advances in Neural Information Processing Systems}, vol.~35, pp. 24\,824--24\,837, 2022.

\bibitem{mernik2005and}
M.~Mernik, J.~Heering, and A.~M. Sloane, ``When and how to develop domain-specific languages,'' \emph{ACM computing surveys (CSUR)}, vol.~37, no.~4, pp. 316--344, 2005.

\bibitem{noury2020deep}
Z.~Noury and M.~Rezaei, ``Deep-captcha: a deep learning based captcha solver for vulnerability assessment,'' \emph{arXiv preprint arXiv:2006.08296}, 2020.

\bibitem{goodfellow2014explaining}
I.~J. Goodfellow, J.~Shlens, and C.~Szegedy, ``Explaining and harnessing adversarial examples,'' \emph{arXiv preprint arXiv:1412.6572}, 2014.

\bibitem{captcha-background}
R.~Gossweiler, M.~Kamvar, and S.~Baluja, ``What's up captcha? a captcha based on image orientation,'' in \emph{Proceedings of the 18th international conference on World wide web}, 2009, pp. 841--850.

\bibitem{captcha-survey}
V.~P. Singh and P.~Pal, ``Survey of different types of captcha,'' \emph{International Journal of Computer Science and Information Technologies}, vol.~5, no.~2, pp. 2242--2245, 2014.

\bibitem{captcha-solver-1}
G.~Ye, Z.~Tang, D.~Fang, Z.~Zhu, Y.~Feng, P.~Xu, X.~Chen, and Z.~Wang, ``Yet another text captcha solver: A generative adversarial network based approach,'' in \emph{Proceedings of the 2018 ACM SIGSAC conference on computer and communications security}, 2018, pp. 332--348.

\bibitem{captcha-solver-2}
Z.~Noury and M.~Rezaei, ``Deep-captcha: a deep learning based captcha solver for vulnerability assessment,'' \emph{arXiv preprint arXiv:2006.08296}, 2020.

\bibitem{captcha-solver-3}
M.~Motoyama, K.~Levchenko, C.~Kanich, D.~McCoy, G.~M. Voelker, and S.~Savage, ``Re:$\{$CAPTCHAs—Understanding$\}$$\{$CAPTCHA-Solving$\}$ services in an economic context,'' in \emph{19th USENIX Security Symposium (USENIX Security 10)}, 2010.

\bibitem{captcha-solver-4}
M.~Korakakis, E.~Magkos, and P.~Mylonas, ``Automated captcha solving: An empirical comparison of selected techniques,'' in \emph{2014 9th International Workshop on Semantic and Social Media Adaptation and Personalization}.\hskip 1em plus 0.5em minus 0.4em\relax IEEE, 2014, pp. 44--47.

\bibitem{llm-survey}
Y.~Chang, X.~Wang, J.~Wang, Y.~Wu, L.~Yang, K.~Zhu, H.~Chen, X.~Yi, C.~Wang, Y.~Wang \emph{et~al.}, ``A survey on evaluation of large language models,'' \emph{ACM Transactions on Intelligent Systems and Technology}, 2023.

\bibitem{deng2023pentestgpt}
G.~Deng, Y.~Liu, V.~Mayoral-Vilches, P.~Liu, Y.~Li, Y.~Xu, T.~Zhang, Y.~Liu, M.~Pinzger, and S.~Rass, ``Pentestgpt: An llm-empowered automatic penetration testing tool,'' 2023.

\bibitem{meng2024large}
R.~Meng, M.~Mirchev, M.~B{\"o}hme, and A.~Roychoudhury, ``Large language model guided protocol fuzzing,'' in \emph{Proceedings of the 31st Annual Network and Distributed System Security Symposium (NDSS)}, 2024.

\bibitem{wang2023bot}
H.~Wang, X.~Luo, W.~Wang, and X.~Yan, ``Bot or human? detecting chatgpt imposters with a single question,'' \emph{arXiv preprint arXiv:2305.06424}, 2023.

\bibitem{gao2021captcha}
\BIBentryALTinterwordspacing
Y.~Gao, H.~Gao, S.~luo, Y.~Zi, S.~Zhang, W.~Mao, P.~Wang, Y.~Shen, and J.~Yan, ``Research on the security of visual reasoning {CAPTCHA},'' in \emph{30th USENIX Security Symposium (USENIX Security 21)}.\hskip 1em plus 0.5em minus 0.4em\relax USENIX Association, Aug. 2021, pp. 3291--3308. [Online]. Available: \url{https://www.usenix.org/conference/usenixsecurity21/presentation/gao}
\BIBentrySTDinterwordspacing

\bibitem{dinh2023recent}
N.~T. Dinh and V.~T. Hoang, ``Recent advances of captcha security analysis: a short literature review,'' \emph{Procedia Computer Science}, vol. 218, pp. 2550--2562, 2023.

\bibitem{andrew2023captcha}
\BIBentryALTinterwordspacing
A.~Searles, Y.~Nakatsuka, E.~Ozturk, A.~Paverd, G.~Tsudik, and A.~Enkoji, ``An empirical study \& evaluation of modern {CAPTCHAs},'' in \emph{32nd USENIX Security Symposium (USENIX Security 23)}.\hskip 1em plus 0.5em minus 0.4em\relax Anaheim, CA: USENIX Association, Aug. 2023, pp. 3081--3097. [Online]. Available: \url{https://www.usenix.org/conference/usenixsecurity23/presentation/searles}
\BIBentrySTDinterwordspacing

\bibitem{kumar2022systematic}
M.~Kumar, M.~Jindal, and M.~Kumar, ``A systematic survey on captcha recognition: types, creation and breaking techniques,'' \emph{Archives of Computational Methods in Engineering}, vol.~29, no.~2, pp. 1107--1136, 2022.

\bibitem{291062}
\BIBentryALTinterwordspacing
A.~Searles, Y.~Nakatsuka, E.~Ozturk, A.~Paverd, G.~Tsudik, and A.~Enkoji, ``An empirical study \& evaluation of modern {CAPTCHAs},'' in \emph{32nd USENIX Security Symposium (USENIX Security 23)}.\hskip 1em plus 0.5em minus 0.4em\relax Anaheim, CA: USENIX Association, Aug. 2023, pp. 3081--3097. [Online]. Available: \url{https://www.usenix.org/conference/usenixsecurity23/presentation/searles}
\BIBentrySTDinterwordspacing

\bibitem{patent}
CAPTCHA\_Patents, \url{https://patents.google.com/?q=(CAPTCHA+)&oq=CAPTCHA}.

\bibitem{wei2023chainofthought}
J.~Wei, X.~Wang, D.~Schuurmans, M.~Bosma, B.~Ichter, F.~Xia, E.~Chi, Q.~Le, and D.~Zhou, ``Chain-of-thought prompting elicits reasoning in large language models,'' 2023.

\bibitem{chen2023theoremqa}
W.~Chen, M.~Yin, M.~Ku, P.~Lu, Y.~Wan, X.~Ma, J.~Xu, X.~Wang, and T.~Xia, ``Theoremqa: A theorem-driven question answering dataset,'' 2023.

\bibitem{gpt4-p}
gpt-4-vision preview, \url{https://platform.openai.com/docs/guides/vision}.

\bibitem{gemini-p}
gemini-pro vision, \url{https://labelbox.com/product/model/foundry-models/google-gemini-pro-vision/}.

\bibitem{vaswani2023attention}
A.~Vaswani, N.~Shazeer, N.~Parmar, J.~Uszkoreit, L.~Jones, A.~N. Gomez, L.~Kaiser, and I.~Polosukhin, ``Attention is all you need,'' 2023.

\bibitem{yang2023chatgpt}
L.~Yang, H.~Chen, Z.~Li, X.~Ding, and X.~Wu, ``Chatgpt is not enough: Enhancing large language models with knowledge graphs for fact-aware language modeling,'' 2023.

\bibitem{sql}
SQL, \url{https://en.wikipedia.org/wiki/SQL}.

\bibitem{3vl}
Three-valued\_logic, \url{https://en.wikipedia.org/wiki/Three-valued_logic}.

\bibitem{ftllama}
F.~tuned~code llama, \url{https://github.com/ragntune/code-llama-finetune/blob/main/fine-tune-code-llama.ipynb}.

\bibitem{microsoft-hips}
K.~Chellapilla, K.~Larson, P.~Simard, and M.~Czerwinski, ``Designing human friendly human interaction proofs (hips,'' 04 2005, pp. 711--720.

\bibitem{von2003captcha}
L.~Von~Ahn, M.~Blum, N.~J. Hopper, and J.~Langford, ``Captcha: Using hard ai problems for security,'' in \emph{Advances in Cryptology—EUROCRYPT 2003: International Conference on the Theory and Applications of Cryptographic Techniques, Warsaw, Poland, May 4--8, 2003 Proceedings 22}.\hskip 1em plus 0.5em minus 0.4em\relax Springer, 2003, pp. 294--311.

\bibitem{minigpt4}
miniGPT 4, \url{https://minigpt-4.github.io/}.

\bibitem{arkose}
Arkose\_Labs, \url{https://www.arkoselabs.com/}.

\bibitem{geetest}
Geetest, \url{https://www.geetest.com/en/}.

\bibitem{netease}
NetEase\_Yidun, \url{https://dun.163.com/locale/en}.

\bibitem{zhao2023large}
\BIBentryALTinterwordspacing
Z.~Zhao, W.~S. Lee, and D.~Hsu, ``Large language models as commonsense knowledge for large-scale task planning,'' in \emph{RSS 2023 Workshop on Learning for Task and Motion Planning}, 2023. [Online]. Available: \url{https://openreview.net/forum?id=tED747HURfX}
\BIBentrySTDinterwordspacing

\end{thebibliography}
